\documentclass[12pt]{article}
\usepackage[utf8]{inputenc}
\usepackage{fullpage}
\usepackage{amssymb, amsmath, amsthm, amsfonts}
\usepackage{authblk}
\usepackage{graphicx}
\usepackage{color}
\usepackage{mathrsfs}
\usepackage{physics}
\usepackage{mathtools}
\usepackage{tikz}
\usepackage{faktor}
\usepackage{tikz-cd}
\usepackage{bm}
\usepackage[hidelinks]{hyperref}
\usepackage{enumitem}
\usepackage[mathlines]{lineno}
\usepackage{tikz-cd}
\usepackage{bm}
\usepackage{csquotes}
\usepackage{float}

\usepackage[UKenglish]{isodate}
\usepackage[UKenglish]{babel}

\usepackage[backend=bibtex, style=vancouver, sorting=none]{biblatex}
\renewbibmacro{in:}{%
  \ifentrytype{article}{}{\printtext{\bibstring{in}\intitlepunct}}}

\usepackage{todonotes}

\newcommand{\R}{\mathbb{R}}

\DeclareMathOperator*{\argmax}{\text{arg\,max}}

\DeclareMathOperator{\E}{\mathbb{E}}
\renewcommand{\grad}{\nabla}

\usepackage{subcaption}
\title{On Bayesian Mechanics: A Physics of and by Beliefs}

\author[1,2,*]{Maxwell J. D. Ramstead}
\author[1,3,4,5,*]{Dalton A. R. Sakthivadivel}
\author[1,6,7,8]{Conor Heins}
\author[1,9]{Magnus Koudahl}
\author[1,10]{Beren Millidge}
\author[2,11]{Lancelot Da Costa}
\author[1,12]{Brennan Klein}
\author[1,2]{Karl J. Friston}
\affil[1]{VERSES Research Lab, Los Angeles, California, 90016, USA}
\affil[2]{Wellcome Centre for Human Neuroimaging, University College London,\protect\\London WC1N 3AR, UK}
\affil[3]{Department of Mathematics, Stony Brook University, Stony Brook, New York, USA}
\affil[4]{Department of Physics and Astronomy, Stony Brook University, Stony Brook, New York, USA}
\affil[5]{Department of Biomedical Engineering, Stony Brook University, Stony Brook, New York, USA}
\affil[6]{Department of Collective Behaviour, Max Planck Institute of Animal Behavior,\protect\\78464 Konstanz, Germany}
\affil[7]{Department of Biology, University of Konstanz, 78464 Konstanz, Germany}
\affil[8]{Centre for the Advanced Study of Collective Behaviour, University of Konstanz,\protect\\78464 Konstanz, Germany}
\affil[9]{Department of Electrical Engineering, Eindhoven University of Technology,\protect\\Eindhoven, The Netherlands}
\affil[10]{Brain Network Dynamics Unit, University of Oxford, Oxford, UK}
\affil[11]{Department of Mathematics, Imperial College London, London SW7 2AZ, UK}
\affil[12]{Network Science Institute, Northeastern University, Boston, Massachusetts, USA}
\affil[*]{\emph{These authors contributed equally and are listed alphabetically by surname.}}

\date{\today}

\begin{document}

\maketitle

\begin{abstract}
    
The aim of this paper is to introduce a field of study that has emerged over the last decade called Bayesian mechanics. Bayesian mechanics is a probabilistic mechanics, comprising tools that enable us to model systems endowed with a particular partition (i.e., into particles), where the internal states (or the trajectories of internal states) of a particular system encode the parameters of beliefs about external states (or their trajectories). These tools allow us to write down mechanical theories for systems that look as if they are estimating posterior probability distributions over the causes of their sensory states. This provides a formal language for modelling the constraints, forces, potentials, and other quantities determining the dynamics of such systems, especially as they entail dynamics on a space of beliefs (i.e., on a statistical manifold). Here, we will review the state of the art in the literature on the free energy principle, distinguishing between three ways in which Bayesian mechanics has been applied to particular systems (i.e., path-tracking, mode-tracking, and mode-matching). We go on to examine a duality between the free energy principle and the constrained maximum entropy principle, both of which lie at the heart of Bayesian mechanics, and discuss its implications.   

\end{abstract}

\vskip0.2cm

\thanks{{\footnotesize \parbox{\linewidth-24pt}{Corresponding author: Dalton A. R. Sakthivadivel (dalton.sakthivadivel@\{stonybrook.edu, verses.io\}, \url{https://darsakthi.github.io}), ORCiD 0000-0002-7907-7611. \vskip0.2cm Keywords: free energy principle, active inference, Bayesian mechanics, information geometry, maximum entropy, gauge theory.}}}

\tableofcontents

\section{Introduction}\label{sec:intro}

In this paper, we aim to introduce a field of study that has begun to emerge and consolidate over the last decade\textemdash called Bayesian mechanics\textemdash which might provide the first steps towards a general mechanics of self-organising and complex adaptive systems \cite{Friston2019, DaCosta2021, Friston2021a, Ueltzhoffer2020, Sakthivadivel2022a, Sakthivadivel2022b}. Bayesian mechanics involves modelling physical systems that look as if they encode probabilistic beliefs about the environment in which they are embedded, and in particular, about the ways in which they are coupled to that environment. Bayesian mechanics thereby purports to provide a mathematically principled explanation of the striking property of all things that exist over some time period, namely: that they come to acquire the statistics of their embedding environment, and seem thereby to encode a probabilistic representation of that environment \cite{RN1250, RN1096}. Bayesian mechanics is premised on the idea that the physical mechanics of particular kinds of systems are systematically related to a mechanics of information, or the mechanics of the probabilistic beliefs that such systems encode. Bayesian mechanics describes physical systems in terms of a pair of complementary spaces linked by certain laws of motion: the space of probability distributions over the physical states of a system (the beliefs of an observer about its environment, say), and a simultaneous space of probability distributions encoded or entailed by the system, which are linked by approximate Bayesian inference. Bayesian mechanics is premised on the conjugation of the \textit{dynamics of the beliefs} of a system (i.e., their time evolution in a space of beliefs) and the \textit{physical dynamics} of the system encoding those beliefs (i.e., their time evolution in a space of possible states of trajectories) \cite{Sakthivadivel2022b, DaCosta2021}; the resulting mathematical structure is known as a ``conjugate information geometry'' in \cite{Friston2019}, where one should note that ``conjugate'' is a synonym for ``adjoint'' or ``dual''). Using the tools of Bayesian mechanics, we can form mechanical theories for a self-organising system that looks as if it is modelling its embedding environment. Thus, Bayesian mechanics describes the image of a physical system as a flow in a conjugate space of probabilistic beliefs held by the system, and describes the systematic relationships between both perspectives.

It is often said that systems that are able to preserve their organisation over time, such as living systems, appear to resist the entropic decay and dissipation that are dictated by the second law of thermodynamics (this view is often attributed to Schr\"odinger, \cite{Schrodinger1944}). This is, in fact, untrue, and something of a sleight of hand, as Schr\"odinger himself knew well: self-organising systems, and living systems in particular, not only conform to the second law of thermodynamics, which states that the internal entropy of an isolated system always increases, but conform to it exceptionally well\textemdash and in doing so, they also maintain their structural integrity \cite{Schrodinger1944, morowitz2007energy, still2012thermodynamics, perunov2016statistical, jeffery2019statistical, Ueltzhoffer2020, Ueltzhoffer2021}. The foundations of Bayesian mechanics have been laid out by the pioneers of the physics of complex adaptive systems and of the study of natural and artificial intelligence. Bayesian mechanics builds on these foundational methods and tools, which have been applied to develop mathematical theories and computational models, allowing us to study the seemingly paradoxical emergence of stable structure as a special case of entropic dissipation \cite{Nicolis1977, Prigogine1978, England2015}. Bayesian mechanics has origins in variational principles from other fields of physics and statistics, such as Jaynes' principle of maximum entropy \cite{Jaynes1957} and the principle of stationary action, and draws on a broad, multidisciplinary array of results from information theory and geometry \cite{Mackay2003, amari, parr-info-geo}, cybernetics and artificial intelligence \cite{Barlow1961, Barlow1974}, computational neuroscience \cite{Edelman2001, Montague2012}, gauge theories for statistical inference and statistical physics \cite{polettini, Friston2015, Sengupta2016, Sengupta2017, Sakthivadivel2022b}, as well as stochastic thermodynamics and non-equilibrium physics \cite{seifert_stochastic_2012, England2015, parr_markov_2020}. Bayesian mechanics builds on these tools and technologies, allowing us to write down mechanical theories for the particular class of physical systems that look as if as if they are estimating posterior probability densities over (i.e., estimating and updating their beliefs about) the causes of their observations.

In this paper, we discuss the relationship between \textit{dynamics}, \textit{mechanics}, and \textit{principles}. In physics, the ``dynamics" of a system usually refers to descriptions (i.e., phenomenological accounts) of how something behaves: dynamics tell us about changes in position and the forces that cause such changes. Dynamics are descriptive, but are not necessarily explanatory: they are not always directly premised on things like laws of motions. We move \textit{from description to explanation} via mechanics, or mechanical theories: specific mathematical theories formulated to explain where dynamics come from, by providing a formulation of the relationships between change, movement, energy (or force), and position. Finally, principles are prescriptive: they are compact mathematical statements in light of which mechanical theories can be interpreted. That is, if a mechanical theory explains \textit{how} a system behaves the way that it does, a principle explains \textit{why}. For instance, classical mechanics provides us with equations of motion to explain \textit{how} the dynamics of non-relativistic bodies are generated, relating the changes in position of the system to its potential and kinetic energies; whilst the principle of stationary action tells us \textit{why} such relations obtain, i.e., the real path of the system is the one where the accumulated difference between these two energies is at a minimum. Likewise, Bayesian mechanics is a set of mechanical theories designed to explain the dynamics of systems that look as if they are driven by probabilistic beliefs about an embedding environment.

We have said that mechanics rest on prescriptive principles. At the core of Bayesian mechanics is the variational free energy principle (FEP). The FEP is a mathematical statement that says something fundamental (i.e., from first principles) about what it means for a system to exist, and to be ``the kind of thing that it is.'' The FEP provides an interpretation of mechanical theories for systems that look as if they have beliefs. The FEP is thereby purported to explain why self-organising systems seem to resist the local tendency to entropic decay, by acting to preserve their structure. The FEP builds on decades of previous work redefining classical and statistical mechanics in terms in terms of surprisal and entropy (e.g., the pioneering work of \cite{Prigogine1978, Nicolis1977, England2015}). Surprisal is defined as the log-probability of an event: heuristically, it quantifies how implausible a given state of a process or outcome of a measurement is, where high surprisal is associated to states or outcomes with a low probability of being observed (in other words, those states something would not, typically, be found in).\footnote{We mean ``typical" in a twofold statistical sense of an event which is typical of some probability density, i.e., both (i) in the sense of a state for which some sample is unsurprising, and (ii) in the sense of a quantity which is characteristic of the ensemble limit of that system (that is, a state which the system is likely to evolve to asymptotically). This is known to concentrate paths-based formalisms into state-based formalisms. See \cite{touchette} and especially \cite{ying-jen} for important remarks on this.} Entropy is the expected or average surprisal of states or outcomes. It is also a measure of the spread of some probability distribution or density, and quantifies the average information content of that distribution \cite{shannon1948mathematical}. Variational free energy is a tractable (i.e., computable) upper bound on surprisal; negative free energy is known as the evidence lower bound or ELBO in machine learning \cite{blei2017variational}. The FEP describes self-organization as a flow towards a free energy minimum. It has been known that one can use the FEP to write down the flow of dynamical systems as self-organising by avoiding surprising exchanges with the environment and thereby minimizing entropic dissipation over time; e.g., \cite{Friston2019, fields2021free}. The FEP consolidates this into a modelling method, analogous to the principles of maximum entropy or stationary action. That is, the FEP is not a metaphysical statement about what things ``really are." Rather, the FEP starts from a stipulative, particular definition of what it means to be a thing, and then can be used to write down mechanical theories for systems that conform to this definition of thing-ness \cite{Friston2019, Friston2021a}.

Before proceeding, we highlight the distinction between two meanings of the word ``belief'': a probabilistic one, where the term ``belief'' is used in the technical sense of Bayesian statistics, to denote a probability density over some support, and thereby formalises a belief of sorts about that support; and a propositional or folk understanding of the term, common in philosophy and cognitive science, which entails some kind of semantic content with verification  conditions (e.g., truth-conditional ones). In this paper, we always mean the former, probabilistic sense of ``belief''; and we will use the terms ``belief'' and ``probability density'' interchangeably. 

With that caveat in place, Bayesian mechanics is specialised for particular systems that have a partition of states, with one subset parameterising probability distributions or densities over another. Bayesian mechanics articulates mathematically a precise set of conditions under which physical systems can be thought of as being endowed with probabilistic (conditional or Bayesian) beliefs about their embedding environment. Formally, Bayesian mechanics is about so-called ``particular systems'' that are endowed with a ``particular partition'' \cite{Friston2019}\textemdash i.e., into particles, which are coupled to, but separable from, their embedding environment. By ``particular system," we mean a system that has a specific (i.e., ``particular") partition into internal states, external states, and intervening blanket states, which instantiate the coupling between inside and outside (the ``Markov blanket"). The internal and blanket states can then be cast as constituting a ``particle," hence the name of the partition.\footnote{In the FEP literature, the word ``system'' usually refers to the set of coupled stochastic differential equations that cover both a particle (internal and blanket states or paths) and its embedding external environment (external states or paths). In other words, the system of concern in the FEP formulation is not only the particle or self-organising system, but rather the dynamics of coupled particle-environment system. (This is sometimes discussed in terms of an extended brain-body-environment system \cite{clark2015surfing}, or an agent-environment loop \cite{Sakthivadivel2022b}.) This is why the word ``particle'' was introduced in this literature: to ensure there is no ambiguity about what was meant by a ``system.'' We note that there are some inconsistencies in the literature, owing to the fact that these terms are not used homogeneously across physics.} Under the FEP, the internal states of a physical system can be modelled as encoding the parameters of probabilistic beliefs, which are (probability density) functions whose domain are quantities that characterise the system (e.g. states, flows, trajectories, other measures).

In a nutshell, Bayesian mechanics is set of a physical, mechanical theories about the beliefs encoded or embodied by internal states and how those beliefs evolve over time: it provides a formal language to model the constraints, forces, fields, manifolds, and potentials that determine how the internal states of such systems move in a space of beliefs (i.e., on a statistical manifold). Because these probabilistic beliefs depend on parameters that are physically encoded by the internal states of a particle, the resulting statistical manifolds (or belief spaces) and the flows along them have a non-trivial, systematic relationship to the physics of the system that sustain them. This is operationalised by applying the FEP: we model the behaviour of a particular system by a path of stationary action over free energy, given a function (called a \textit{synchronization map}) that defines the manner in which internal and external states are synchronized across the boundary (or Markov blanket) that partitions any such dynamical system (should that partition exist). In summary, Bayesian mechanics is about the image of a physical system in the space of beliefs, and the connections between these representations: that is, it takes the internal states of a particular system (and their dynamics) and maps them into a space of probability distributions (and trajectories or paths in that space), and vice versa.

Two related mathematical objects form part of the core of the FEP, and will play a key role in our account of Bayesian mechanics: (i) ontological potentials or constraints, and (ii) the mechanics of systems that are driven by such potentials. An ontological potential, on this account, is similar to other potentials in physics, e.g., gravitational or electromagnetic potentials. It is a scalar quantity that defines an energy landscape, the gradients of which determine the forces (vector fields) to which the system is subject. Such potentials are \textit{ontological} because they characterises \textit{what it is to be} the kind of thing that a thing is: they allow us to specify equations of motion that a system must satisfy to remain the kind of thing that it is. 

Ontological potentials or constraints provide a mathematical definition of what it means for a particular system to be the kind of system that it is: they enable us to specify the equations of motion of particular systems (i.e, their characteristic paths through state space, the way that they evolve over time, the states that they visit most often, etc), based on a description of what sorts of states or paths are typical of that kind of system. We review these notions in technical detail in Sections \ref{sec:sentientsystems} and \ref{sec:math_prelim}. In particular, Bayesian mechanics is concerned with the relationship between the ontological potentials or constraints, and the flows, paths, and manifolds that characterise the time evolution of systems with such potentials, allowing us to stake out a new view on adaptive self-organisation in physical systems. 

We shall see that this description via the FEP always comes with a dual, or complementary, perspective on the same dynamics, which is derived from maximum entropy. This view is about the probability density that the system samples under, and how that density is enforced or evolves over time.\footnote{We are using the words ``dual," ``dualisation," and ``dually" in a semi-technical sense; see Section \ref{sec:maxent}. Briefly, dual objects are precise opposites of one another. The duality of two maps or objects, which we call an adjoint pair, means that they share intrinsic features, but exhibit relationships to other objects in opposite directions. The reader should note that the term adjoint pair is most often seen in category theory, but that we do not explicitly consider category-theoretic notions here.} We consider at length the duality between the FEP and the constrained maximum entropy principle (CMEP), showing that they are two perspectives on the same thing. This provides a unifying point of view on adaptive, self-organising dynamics, which embraces the duality of perspectives: that of adaptive systems on their environment (and themselves), and that of the ambient heat bath, in which they are embedded (and into which all organised things ultimately decay). 

These points of view might seem opposed, at least \textit{prima facie}: after all, persistent, complex adaptive systems appear organised to resist entropic decay and dissipation; while the inevitable, existential thermodynamic imperative of all organised things embedded in the heat bath is to dissipate into it \cite{Schrodinger1944}. The resolution of this apparent tension is a core motivation for dualising the entire construction. Just as we can think of controlled systems that maintain their states around characteristic, unsurprising set-points \cite{DaCosta2021}\textemdash despite perturbations from the environment\textemdash we can view a self-organising system as a persistent, cohesive locus of states that is embedded within an environment, and which is countering the tendency of the environment to dissipate it. This ``agent-environment'' or ``relational'' symmetry is fundamental to almost all formal approaches to complex systems, which are rooted in the interactions between open systems \cite{rosen1, rosen2, tony, hedges, toby, matteo}, making it an attractive framework for understanding complexity. 

In particular, self-organisation can be viewed in two ways. One is from the perspective of the ``self,'' inhabiting the point of view of the individuated thing that is distinct from other things in its environment. From this perspective, provided by the FEP, one can ask how particular systems interpret their environment and maintain their ``selves''\textemdash the kinds of structure typical of the kind of thing that they are. This requires engaging in inference about the causes of internal or sensory states. Dually, one can view this from the perspective of ``organisation''\textemdash i.e., from the outside peering in, modelling what it looks like for a structure to remain cohesive, and not dissipate into its environment over some appreciable timescale. This latter perspective is like asking about the internal \textit{states} of some system, rather than the \textit{beliefs carried} by internal states (as one might under an FEP-theoretic lens). That both stories concern the self-organisation of the system, but model it in different ways, is no accident.

In the same, dual sense, asking about organisation is like being an observer or a modeller situated in the external world, formulating beliefs about the internal states of a particular system. These viewpoints are equivalent, in that they tell the same story about inference and the dynamics of self-organisation. This duality allows us to view the FEP and Bayesian mechanics through a number of complementary lenses. The advantage to changing our perspective is that we can compare the FEP to the view from maximum entropy, which is more familiar in standard mathematics and physics. In particular, it should provide us with a systematic method to relate the dynamics and mechanics of organised systems to the dynamics and mechanics of the beliefs encoded or embodied by these organised systems, recovering fundamental predecessor formulations to Bayesian mechanics and the FEP in the language of the physics of self-organisation.

The argumentative sequence of this paper is as follows. The overall paper is divided into three main parts. The first part of the paper is written as a (relatively) reader-friendly, high-level descriptive summary of the free energy principle literature, which spans nearly two decades of development. We first offer some preliminary material on dynamics, mechanics, field theories, and principles, and provide some motivation for the emergence of Bayesian mechanics. We then discuss the state of the art of Bayesian mechanics in quite some depth. We provide a narrative review the core formalisms and results of Bayesian mechanics. We comprehensively review the FEP as it appears in the literature, and distinguish three main ways in which it has been applied to model the dynamics of particular systems. We call these \textit{path-tracking}, \textit{mode-tracking}, and \textit{mode-matching}. The second part of the paper introduces, again at a high level, a new set of results that have become available only recently, which concern the duality between the free energy principle and the principle of maximum entropy subject to a particular constraint; and which are more mathematically involved, drawing in particular on gauge theory. We make a short detour to discuss gauge theories, maximum entropy, and dualisation. With this in place, we examine the duality of the FEP and the CMEP. The final part of the paper discusses the burgeoning philosophy of Bayesian mechanics. We discuss the implications of the duality between the FEP and the CMEP for Bayesian mechanics, and sketch some directions for future work. Finally, taking stock, we chart a path to a generalisation of this duality to even more complex systems, allowing for the systematic study of systems that exist far from equilibrium and elude steady state densities or stationary statistics\textemdash an area of study that we designate as $G$-theory, covering the duality of Bayesian mechanics of paths and entropy on paths (or calibre), and beyond.

The reader should note that this paper is not a standalone treatment of the Bayesian mechanics and the free energy principle, and should instead be seen as a more conceptually-oriented companion paper to the technical material that is reviewed; as such, we will often opt for qualitative descriptions instead of explicit equations, and refer the reader to the technical material for detailed examinations of assumptions and proofs. It should be noted that the fields encompassing Bayesian mechanics, the free energy principle, and the maximum entropy principle are inherently technical ones that presuppose and leverage detailed formal constructs and concepts. We aim for this paper to be relatively self-contained, and provide some introductory material to facilitate reading; but we assume that the reader has a working knowledge of dynamical systems theory (specifically, the state or phase space formalism), calculus (especially ordinary and stochastic differential equations), and probability or information theory. A familiarity with gauge theory is also useful in reading the second main part of the paper. The philosophical denouement of the paper should be accessible to readers with relatively little background in mathematics and physics.
\section{An overview of the idea of mechanics}\label{sec:preliminaries}

Before diving into Bayesian mechanics proper, we begin by reviewing some of the core concepts that underwrite contemporary theoretical physics. 

In formal approaches to the study of physics, a description of the behaviour of a specific object is the very bottom of a hierarchy of theory-building. As discussed in the introduction, the dynamics of a system constitute a description of the forces to which that system is subject, which is typically specified via laws or equations of motion (i.e, a mechanics). Before we can derive a mathematical description of the behaviour of something, we need a large amount of other information accounting for where those equations of motion come from. A mechanical theory, for instance, is a mathematical theory that tells us how force, movement, change, and position relate to each other. In other words, mechanical theories tell us how a thing should behave; and, given some specific system, we can use the mechanical theory to specify its dynamics. This distinction between a phenomenological (or merely descriptive) model and a physical, mechanical theory usually lies in that mechanical theories can be derived from an underlying principle, like the stationary action principle. Thus, the resultant mechanical theory specifies precisely what systems that obey that principle do\textemdash and conversely, the principle provides an interpretation of that mechanical theory, given a set of system-level details relevant for the sought dynamical picture. 

The word ``theory'' is polysemous. A quick overview of the key notions in the philosophy of scientific modelling is useful to clarify what we mean here (see \cite{Andrews2021} for an excellent overview; also see \cite{andrews2022making, weisberg2007modeler, weisberg2012simulation}). What we have called ``dynamics,'' ``mechanics,'' and ``principles'' are, ultimately, mathematical structures (in mathematics, these are also called mathematical theories). The \textit{content} of a mathematical theory or structure is \textit{purely formal}: e.g., clearly, the axioms and theorems of calculus and probability theory are not intrinsically \textit{about} any real empirical thing in particular. What are usually called ``scientific theories'' or ``empirical theories'' comprise a mathematical structure and what might be called an \textit{empirical application}, interpretation, or construal of that structure, which relates the constructs of the mathematical structure to things in the world, e.g., to specific observable features of systems that exist.\footnote{To avoid ambiguity, we will reserve the term ``interpretation'' for the way that a principle makes sense of \textit{why} a mechanical theory works the way that it does; and reserve the term ``empirical application'' to refer to the systematic association of some aspects of a mathematical theory or structure to aspects of the empirical world. See \cite{Andrews2021, andrews2022making} for a closely related discussion.} 

It is sometimes said that principles in physics, such as the principle of stationary action, are \textit{not falsifiable}, strictly speaking; and this, despite playing an obviously prominent role in scientific investigation, which is ultimately grounded in empirical validation (at least \textit{prima facie}). We can make sense of this in light of the distinction between mathematical theories (i.e., what we have called mechanical theories and principles) and their empirical applications. The resolution of the tension lies in noting that absent some specific empirical application, mathematical structures are not meant to say anything specific about empirical phenomena. Indeed, as \cite{andrews2022making} and \cite[see Remark 5.1]{Sakthivadivel2022b} argue, the possibility of introducing productive ``abuses of notation'' by deploying one same mathematical structure to explain quite different phenomena is part of what makes formal modelling such a powerful scientific tool in the first place. 

We have said that dynamics are descriptive, but that they are not necessarily explanatory. There is a long tradition of argument according to which dynamical systems models are not inherently explanatory (e.g., \cite{van1995dynamical, van1998dynamical, kaplan2011explanatory}), since they do not necessarily appeal to an explanatory mechanism, and instead provide a convenient formal summary of behaviour. This is the main difference between, for example, Kepler's account of the motion of celestial bodies, which was merely descriptive (and so, a dynamics according to our definition), and Newton's universal laws of motion, which provide us with a mechanics, apt to explain these dynamics. That is, Kepler's laws of planetary motion are not really equations of motion in the contemporary sense; they are description of heliocentric orbits as elliptical trajectories, and do not provide an explanation of the shape of these orbits (e.g., in terms of things such as mass and gravitational attraction, as Newton would later do; and which we would label as mechanics).

Since the pioneering work of Maxwell, nearly all of modern physics has been formulated in terms of \textit{field theories}. After the turn of the 20th century, all of physics was reformulated in terms of spatially extended fields, due to their descriptive advantages \cite{weinberg}. Fields are a way to formally express how a mechanical theory applies to a system within the confines of a single path on space-time, a so-called world-line. That is, a field constrains equations of motion to apply to specific, physically realizable trajectories in space-time. (Likewise, most of modern physics has been geometrised due to geometry's descriptive advantages \cite{atiyah}. Later, we will see that contemporary physics augments the field theoretical apparatus with the geometric tools of gauge theory.) Mathematically speaking, a field is a $n$-dimensional, abstract object that assigns a value to each point in some space; when that value is a scalar, we call the field a scalar field, of which a special case is a \textit{potential function}. For example, the electromagnetic field assigns a charge density to each point in space (i.e., the electric and magnetic potential energies); the gradient of this potential, in turn, determines the force undergone by a particle in that potential. Similarly, a gravitational field assigns a value to each point in space time, in terms of the work per unit mass that is needed to move a particle away from its inertial trajectory. 

We are concerned here with what we have called Bayesian mechanics. In general, when one speaks about a mechanical theory for some sort of physics\textemdash such as quantum mechanics (a theory describing the behaviour of things at high energies, i.e., very small things moving very quickly, like quantum particles), statistical mechanics (a theory producing the behaviour of systems with probabilistic degrees of freedom, especially the ensemble-level behaviour of large numbers of functionally identical things), or classical mechanics (a theory producing the behaviour of objects and particles in the absence of noise or quantum effects, and at non-relativistic speeds)\textemdash mechanics, which provide the equations of motion that we are interested in, are themselves deduced from some sort of symmetry or optimization principle. Mechanical theories can then be fed data about a \emph{specific} system, like an initial or boundary condition for the evolution of that system, and the equation will return the dynamics of that system.\footnote{Contemporary theoretical physics is premised on the use of optimization principles to determine mechanical theories. The use of optimization principles in the construction of mechanical theories to derive the dynamics of a physical system should not be misinterpreted as being tantamount to claiming that physical systems actually \textit{calculate} their extrema. Principles of symmetry, which are known to underwrite all physics, are something of a scaffolding for the use extremisation\textemdash for instance, on a path of least action, the momentum and energy of free particles is conserved, a time translation symmetry given by Noether's theorem. Optimisation can be seen as a consequence of the desire for symmetries in our picture of the material world. Along the same lines, instead of assuming (indeed, falsely) that physical systems engage in explicit calculation of their dynamics, we only require that there exists a Lagrangian or Lyapunov function for those dynamics: a quantity that varies systematically with those dynamics. Empirically, we know that generic physical systems conform to at least one such symmetry or conservation principle. Indeed, this conformance is a striking \textit{empirical fact} about the universe in which we live, and not merely a mathematical artifact or modelling strategy.}

An example of a mechanical theory operationalising a mathematical principle, which thereby allows us to specify the dynamics of a system, is in the fact that classical objects obey Newton's second law\textemdash i.e., that 
\[
m  \ddot q = -\pdv{q} V(q)
\]
for the position $q$ of some object at a time $t$, and a force on that position, $-\pdv{q} V(q)$. Note that, as discussed above, the force is expressed as the space-derivative $\pdv{q}$ of a gravitational potential, $V(q)$, with respect to position $q$. Newtonian classical mechanics, which is embodied by this equation, gives us the dynamics (i.e., a trajectory) of some classical object when we specify precisely what the potential $V(q)$, initial velocity $\dot q(0)$, initial position $q(0)$, other appropriate boundary conditions, and finally the domain, are. 

\begin{figure}[t!]
    \centering
    \includegraphics[width=0.7\textwidth]{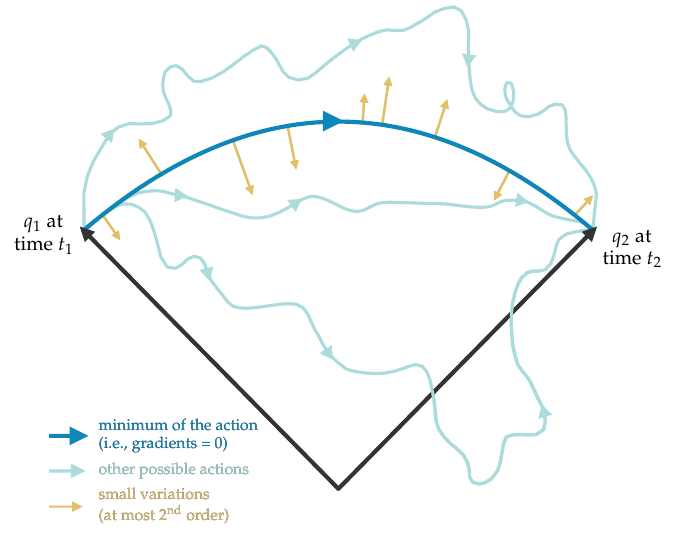}
    \caption{\textbf{Depiction of the principle of stationary action}. This figure shows that the path of least action (darker blue) is the path which is a stationary point of the action\textemdash a path for which the gradient of the action is zero. Here the trajectory is a parabola, like the kind of path one might observe through a gravitational field. On this path, the action changes at most quadratically under the variations in yellow. Other paths in lighter blue are less ``ideal,'' in that they break the precise balance between kinetic and potential energies. That is, these paths do not move in the potential well.}
    \label{fig:least_action}
\end{figure}

If, instead, we had specified \textit{energy functions} for our system based on the force observed, we could have produced the dynamics of this object via Lagrangian mechanics. In classical physics, and more generally, a quantity which summarises the dynamics of a system in terms of the energy on a trajectory through the space of possible states, or path, is called a \textit{Lagrangian} \cite{arnold}. Physically, this is defined in terms of a kinetic energy. We can use a Lagrangian to determine the principle from which classical mechanics follows. Letting $\dot q(t) = v$, Newton's second law comes from the principle of stationary action, since the variation of the following integral of the Lagrangian along a path,
\[
S[q(t)] = \int \frac{mv^2}{2} - V(q) \dd{t},
\]
yields the Euler-Lagrange equation
\[
0 = \frac{m}{2}\pdv{\big(v^2 - V(q)\big)}{q} - \dv{t}\frac{m}{2}\pdv{\big(v^2 - V(q)\big)}{v}
\]
when minimised. The integral mentioned is called an \textit{action functional}, and the variation of the action leads to a rule describing any path of stationary action \cite{feynman}. We can see this in our example: the Euler-Lagrange equation reduces to 
\begin{align*}
0 &= - \pdv{q} V(q) - m \pdv{t} v\\
\implies \\
m \ddot q &= -\pdv{q} V(q)
\end{align*}
after some algebra. This is Newton's second law (noting that the acceleration $a$ is the second derivative of position, i.e., $a$ = $\ddot q$). This result is a summarisation of the fact that systems tend not to use\footnote{Recall, however, that a system is in general unaware of its use of energy; this is a fictive description which makes the stationary action principle more intuitive.} any more energy than they need to accomplish some motion (see \cite{coopersmith} for a pedagogical overview); this translates to the use of the least possible amount of energy to perform a given task. In other words, along the path of stationary action, the change in potential energy is precisely equal to the change in kinetic energy (i.e., their difference is zero), such that no `extra' energy is used, and no `extra' motion is performed. The fact that the accumulated difference in kinetic energy and potential energy is zero reflects this desired law for the exchange of the two quantities\textemdash indeed, this is what underwrites the conservation of energy in physics more generally. It is also what justifies the observation that systems tend to accelerate along forces, and do so by precisely the \textit{force applied}\textemdash no more, and no less. As such, classical mechanics tells us that systems accelerate along forces because they conserve their energy and follow paths of stationary action (i.e., paths over which its variation is zero), and conversely, the theory of classical mechanics comes from the stationary action principle. See Figure \ref{fig:least_action}.

For us, important examples of a principle (with are accompanied by mechanical theories) include the principle of stationary action (which we have just discussed), the maximum entropy principle, and the free energy principle. According to Jaynes, the maximum entropy principle is the principle whereby the mechanics of statistical objects lead to diffusion \cite{villani, jaynes-where, dill}.\footnote{Recall the physical basis for maximum entropy, i.e., that physical systems tend towards the macrostate with the largest number of microstates, given some constraints (or parameters for the assignment of microstates to macrostates, via their probabilities) \cite[Section 3]{Jaynes1957}. It can be mathematically proven that this `spreading' behaviour is what leads to diffusion \cite{villani}.} Likewise, the FEP is the principle according to which organised systems remain organised around system-like states or paths, and the mechanical theory induced by the FEP can be understood as entailing the dynamics required to self-organise. We can understand the former as statistical mechanics\textemdash the behaviour of particles under diffusion\textemdash and we have called the latter Bayesian mechanics. Interestingly, to every physical theory is paired some sort of characteristic geometry, such as symplectic geometry in classical mechanics; moreover, as discussed, mechanical theories are usually seen as the restrictions of field theories to a worldline. By focussing on the aspects of the FEP that relate a physical principle of symmetry to a mechanical theory for the dynamics of some given system, we implicitly introduce notions of geometry and field theory to the FEP, both of which are enormously powerful. We review these ideas in Sections \ref{sec:maxent} and \ref{sec:discussion}.
\section{The free energy principle and Bayesian mechanics: an overview}\label{sec:sentientsystems}

\begin{figure}[t!]
    \centering
    \includegraphics[width=0.9\textwidth]{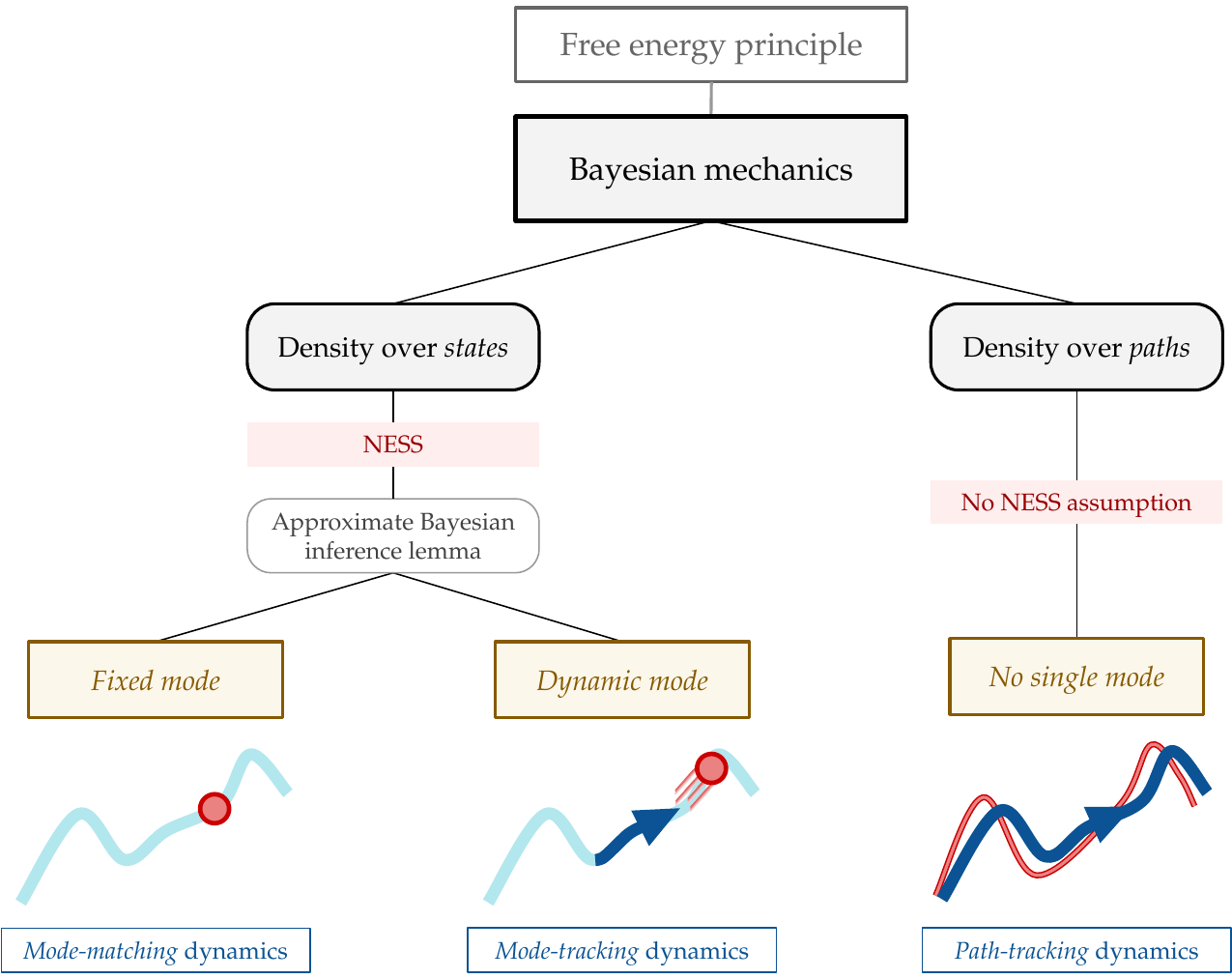}
    \caption{\textbf{Three faces of Bayesian mechanics.} Under the FEP, we can define specific mechanical theories for beliefs, which defines what kinds of self-evidencing are possible in a given regime of mathematical systems. The literature contains three main applications of Bayesian mechanics, which we represent as a tree with two branching points. On the one hand, the FEP has been applied to densities over paths or trajectories of a particular system (the paths-based formulation of FEP, leading to what we call \textit{path-tracking} dynamics) and to densities over states (the density dynamics formulation), which depend on a NESS solution to the mechanics of the system. The density dynamics formulation, in turn, applies to systems with a static mode, and to systems with a dynamic mode; we call the former \textit{mode-matching}, and the latter \textit{mode-tracking}.}
    \label{fig:schematic_three_branches}
\end{figure}

This section reviews key results that have been derived in the literature on the variational free energy principle (FEP), situating them within the broader Bayesian mechanical perspective. We first provide a general introduction to the FEP. We then outline a fairly comprehensive \textit{typology of formal applications} of the FEP that one can find in the literature; these are applications to different kinds of systems with different mathematical properties, which are not often distinguished explicitly. We begin by examining the simplest and most general formulation of the FEP, applied to model \textit{probability densities over paths} of a particular system, often written in generalised coordinates of motion. This general paths-based formulation of the FEP assumes very little about the dynamics of the system, and in particular, does not assume that a non-equilibrium steady state with a well-defined mode exists. We then turn to a formulation of the FEP in terms of the dynamics of a probability density \textit{over states} (i.e., the \textit{density dynamics} formulation of the FEP). This density dynamics formulation, in turn, has taken two main forms in the literature, where the external partition of states has, and does not have, any dynamics to it, respectively. The density dynamics formulation makes stronger assumptions than the paths-based formulation, namely, that the mechanics of the system admit a steady state solution, which allows us to say specific things about the form of the \textit{flow} of a particular system. We discuss a result known as the approximate Bayesian inference lemma that follows from the density dynamics formulation. See Figure \ref{fig:schematic_three_branches}.

\subsection{An introduction to the free energy principle}

In Section \ref{sec:preliminaries}, we said that principles are mathematical theory or structure that are used to write down mechanical theories for a given class of systems. The FEP is precisely such a mathematical principle, which we can leverage to write down mechanical theories for ``things'' or ``particles,'' defined in a particular way. The FEP is the mathematical statement that if something persists over time with a given structure, then it must encode or instantiate a statistical (generative) \textit{model} of its environment. In other words, the FEP tells us that things that maintain their structure in an embedding environment necessarily acquire the statistical structure of that environment. 

Like most of contemporary statistical physics, the FEP starts from probabilistic description of a system\textemdash usually, a system of stochastic differential equations (SDEs). SDEs are used to describe the time evolution or \textit{flow} of a system (i.e., to write down a mechanical theory) in the space of possible states or configurations that it can occupy (what is known as the state or phase space of that system). The SDEs allow us to formulates mechanical theories to explain dynamics with a deterministic component (also known as the \textit{drift} of the SDE) and a stochastic component (the \textit{noise} of the SDE). In the absence of noise, an SDE reduces to an ordinary differential equation, whereby the system evolves deterministically in the direction of the flow.

Formal treatments of the FEP usually begin with a physical system described by an It\^o SDE of the form
\begin{align*}
    \dot{x}(t) = f(x_t) + C \xi (t),
\end{align*}
where $f(x)$ is the drift of the flow of $x$ and $\xi(t)$ is white noise; heuristically, the time derivative of a standard Wiener process $\dd{W_t} / \dd{t}$. The volatility matrix $C$ encodes the spatial directions and magnitude of the noise and yields the diffusion tensor $\Gamma = \frac{C C^{\top}}{2}$, which encodes the covariance of random fluctuations.\footnote{These first few remarks on fluctuations and their variance highlights how the FEP is formulated implicitly in a multi-scale fashion, which has motivated a growing body of work on multi-scale dynamics under the FEP \cite{Heins2022_spins, levin2019computational, RN448}. Indeed, already at this early stage of presentation, the formulation presented here appeals to a separation of timescales: we distinguish between a faster timescale of states or paths that are effectively treated as random fluctuations, and a slower timescale of states or paths that are treated as states or paths \textit{per se}. This is expanded upon and used to effect in \cite{Friston2019}.}

With such a general setup in place, most of contemporary physics, from classical to statistical and quantum mechanics, proceeds to ask whether we can say anything interesting about the probability distribution of different paths or states of the system. The probability distribution over paths or states of a system is usually called a \textit{generative model} in the FEP literature \cite{Ramstead2019enactive}. In statistics, a generative model or probability density is a joint probability density function over some variables. In the FEP literature, the generative model can be analysed in two complementary ways, leading to two main formulations of the FEP: either as a density over states, specifying their probability of being encountered (as opposed to encountering surprising states); or as a density over paths, quantifying the probability of that path (as opposed to other, less probable paths). One can visualize the generative model as a curved (probability density shaped) surface over the space of states or paths of a particular system; the probability of a state of path is associated with the height of the image of the function over the space of states or paths. 

The next step concerns the \textit{particular partition}. We said that the FEP applies to ``things" defined in a particular way, which become models of their embedding environment. Clearly, to talk about ``a model" necessitates a partition into two entities: something that we can identify as the model, and something being modelled. Accordingly, the FEP applies informatively to ``things" defined in a particular way, via a \textit{sparse causal dependence structure} or \textit{sparse coupling}, which is the key construction from which the rest of the FEP follows \cite{ friston2021some, Friston2021a, weak-blankets}. In other words, the FEP is a principle that we can apply to specify the mechanics of systems that have a specific, \textit{particular partition} (i.e., into a thing that is a model and a thing that is being modelled). For such a distinction to hold in physical systems, it must be the case the causal coupling between the thing that instantiates the model and the thing that is modelled evince some kind of sparsity. Consider the informal proof by contradiction: if everything was causally affected by everything else (i.e, if there was no sparse coupling, as in a gas) over some meaningful timescale, then we would not be able to speak of any one ``thing'' as against a backdrop of other things.

So, we partition the entire system $x$ into four components.\footnote{A less traditional but equivalent partition into three parts is also possible, see \cite{barp_geometric_2022}.} Explicitly, we set $x = ( \eta , s,a, \mu)$, where $\mu$ denotes variables pertaining to the model (termed ``internal states'' or ``internal paths''), $\eta$ denotes variables pertaining to the generative process (termed ``external states'' or ``external paths'') and $b = (s, a)$ denotes variables that couple internal and external states\textemdash the \textit{Markov Blanket}\textemdash which, here, comprises sensory and active states (or paths).\footnote{We use the term “Markov blanket” to denote boundary states generally. Most of the time, the relevant boundary states either (i) entail a strict Markov blanket, or else (ii) entail an approximate one. The SCC tells us that we are bound to have a (weak) Markov blanket if the system considered is large enough. We acknowledge that, in some of the more interesting cases, the boundary states are indeed not Markov blankets, strictly speaking. But we have dedicated terminology for those cases: these are, instead, adiabatic or weak blankets, as the case may be; see \cite{weak-blankets}.} Generically speaking, the Markov blanket of the particular partition is precisely the set of degrees of freedom that separate\textemdash but also couple\textemdash one particle (or open system) to another, within a given overarching system \cite{classical-physics}. For instance, in \cite{path-integrals}, the Markov blanket becomes the variable whose space-time path distinguishes one particle from another; this is similar to the argument presented in \cite{virgo2022embracing}. The aim of introducing a particular partition is exactly to introduce the degrees of freedom allowing us to separate one system from another, in such a way that they could in principle engage in inference about (i.e., track) each other. In this sense, the Markov blanket is not special\textemdash it simply entails separability given some variable $b$.\footnote{As we will go on to describe, this serves as an intuition for the commonality of Markov blankets in high-dimensional systems.}

Sensory states are a subset of the Markov blanket: they are those blanket states that are affected by external states and that affect internal states, but that are not affected by internal states. Active states are those blanket states that are affected by internal states and that affect external states, but that are not affected by external states states.\footnote{The Markov blanket condition can also be weakened to a notion which departs from strict conditional independence \cite{weak-blankets}.}

As indicated, this partition assumes \textit{sparse coupling} \cite{heins2022sparse, weak-blankets} among the subset of states or paths (i.e., some subset of the partition evolves independently of another given the dynamics of the blanket), which has the following form:
\begin{align} \label{eq:stoch_blanket}
    f(x) &= 
    \begin{bmatrix}
    f_\eta(\eta,s,a) \\
    f_s(\eta,s,a) \\ 
    f_a(s,a,\mu) \\
    f_\mu(s,a,\mu) 
    \end{bmatrix},\quad 
    C= \begin{bmatrix} 
    C_{\eta} & 0 &0 &0\\
    0 & C_{s} &0&0\\
    0 & 0 & C_{a}&0\\
    0&0&0& C_{\mu}
    \end{bmatrix},
\end{align}
where we use subscripts to denote which subset to which each flow applies. The key point to note is that the flow of internal and active components (i.e., their trajectory through state space) does not depend upon external components (and reciprocally, the flow of external and sensory states or paths does not depend upon internal states or paths). It should be stressed that the blanket is an interface or boundary in \textit{state space}, i.e., it is not necessarily a boundary in spacetime \cite{fields2017eigenforms} (although in some cases, it coincides with one, e.g., the walls of a cell). The internal states (or paths) and their blanket states (or paths) are generally referred to as the \textit{particular states or paths} (i.e., states or paths of a particle); while the internal and active states (or paths) are together called \textit{autonomous states or paths} because they are not influenced by external states (or paths). 

The key point of this construction is that, given such a partition, under the FEP, we can interpret the autonomous partition of the particular system as engaging in a form of Bayesian inference, the exact form of which depends on additional assumptions made about the kind of sparse coupling and conditional independence structures of the particular system being considered \cite{heins2022sparse}. The particular partition is ultimately what licences our analogy with Bayesian inference \emph{per se}, because it licences our interpretation of the internal states of the system as performing (approximate Bayesian or variational) inference. In variational inference, we approximate some ``true'' probability density function $p$ by introducing another probability density called the \textit{variational density} (a.k.a. the recognition density), denoted $q$, with parameters $\mu$. Using variational methods, we vary the parameters $\mu$ until $q$ becomes a good approximation to $p$. In a nutshell, the FEP says that, given a particular partition, the internal states of a particular system encode the sufficient statistics of a variational density over external states (e.g., the mean and precision of a Gaussian probability density). This, as we shall see, induces an internal, statistical manifold in the internal space of states or paths\textemdash and accompanying information geometry.

As stated in Section \ref{sec:intro}, the point of this sort of inference is to minimise the surprisal of particular states or paths of such states. We encountered the idea of minimisation of some action as a principle for mechanics in a more general fashion in Section \ref{sec:preliminaries}. This remains the case here: we can write down the FEP as a ``principle of least surprisal.'' When applied to different kinds of formal systems, we get different types of Bayesian mechanics, in the same sense as we get various sorts of classical mechanics in different mathematical contexts, depending on the assumptions made about the underlying state spaces and action functionals (Newtonian or Lagrangian mechanics, gravitational mechanics, continuum mechanics, and so forth). 

In \cite{friston2022free, path-integrals}, the surprisal on a path (and in particular, a path conditioned on an initial state) 
\[
A[ x(t) ] = -\log p(x(t) \mid x_0)
\]
is suggested as an action for Bayesian mechanics. Here, $x(t)$ is a path of system states at a set of times parameterised by $t$. The path of least action is the expected flow of the system, $f(x_t)$.

A crucial aspect of this formulation is that we can recover Bayesian inference from the minimisation of surprisal. Suppose we apply the surprisal to paths of particular states, i.e., $\pi(t) = (a(t), s(t), \mu(t))$. The mechanics of $\pi(t)$, in terms of the mechanics of the beliefs that $\pi(t)$ holds, is where the Bayesian mechanical story starts. As we just said, the presence of this particular partition implies the existence of a variational density
\begin{equation}\label{recog-density-eq}
q_{\mu}(\eta) = p(\eta \mid \pi)
\end{equation}
at a given time\textemdash in other words, a belief about external states parameterised by internal states. Heuristically, we can think of this as a probabilistic specification of how causes generate consequences. Indeed, we can now say 
\[
-\log p(\pi(t)) = \E_q\big[ \log q(\eta(t)) - \log p(\eta(t)) - \log p(\pi(t) \mid \eta(t)) \big],
\]
and using arguments typical in variational Bayesian inference, this allows us to claim that such systems do engage in inference, via the identity
\begin{equation}\label{kl-div-eq}
-\log p(\pi(t)) = D_\text{KL}[ q(\eta(t)) \| p(\eta(t) \mid \pi(t))]-\log p(\pi(t))
\end{equation}
which holds if and only if the system is inferring the causes of its observations, such that \eqref{recog-density-eq} holds (in which case the KL divergence above is zero). In that sense, any system which minimises its surprisal automatically minimises the variational free energy, licensing an interpretation as approximate Bayesian inference.

In detail\textemdash crucially, this variational free energy is a functional of a probability density over external states or paths that is parameterised by internal states or paths (given some blanket states or paths) and it plays the role of a marginal likelihood or model evidence in Bayesian statistics. This is a key step of the construction, because it connects the entropy \textit{of} the system to the entropy \textit{of its beliefs}, i.e., the entropy of the distribution over internal states $H[p(\mu)]$ and the entropy of the variational (or recognition) density over external states, parameterized by internal states $H[q(\eta)]$. 

\subsection{Applications of the free energy principle to paths, without stationarity or steady state assumptions}

The simplest, most general, and in many ways most natural expression of the FEP, is formulated for the paths of evolution of a particular system \cite{path-integrals}. It is often underappreciated that the FEP was originally formulated in terms of paths, in generalised coordinates of motion \cite{Friston2012, Friston2010}; it is, after all, a way of expressing the principle of stationary action, which tells us about the most likely path that a particle will take under some potential. Much of the mathematics of the FEP goes back to work done in signal processing and Bayesian filtering, which is dynamical, and was developed for neuroimaging \cite{balaji2011bayesian, friston2010generalised}. It would, however, be inaccurate to say that the work on the path integral formulation was abandoned in favour of of other formulations. Indeed, the astute reader will note that the main monograph on the FEP, namely \cite{Friston2019}, was written (albeit sometimes implicitly) in terms of generalised coordinates.

When we operate in paths-based formulation, we operate in what are known as ``generalised coordinates," where the temporal derivatives of a system's flow are considered separately as components of the ``generalised states" of the system \cite{friston2010generalised, kerr2000generalized, path-integrals}. We can use these generalised states to define an instantaneous path in terms of the state space of a system, because they can be read the coefficients of a Taylor expansion of the states as a function of time \cite{friston2022free, path-integrals}. In this formulation, a ``point'' in generalised coordinates corresponds to a possible instantaneous path, i.e., a trajectory over state space or ordered sequence of states; and the FEP, as formulated over paths, concerns probability densities over such instantaneous paths. 

Formally, the paths-based formulation of the FEP says that, for any given path of sensory states, the most likely path of autonomous states (i.e., the path in the joint space of internal and external states) is a minimiser or stationary point of a free energy functional.\footnote{The reader should note that there is a crucial difference between the stationary point of a functional of probability densities (which is also a \textit{probability density}) and the stationary point of the dynamical system \textit{per se}, which is a point in the state space. When we say that a path is a stationary point of a free energy functional, we mean that it is the least surprising path the system can take through state space.} This can be expressed as a variational principle of stationary action, where the action is defined as the \textit{path integral of free energy} \cite{path-integrals}; in the sense that variations of the most likely path do not appreciably change the integral along a trajectory of the free energy $F$, i.e.,
\begin{align*}
    \dot x(t) = f(x_t) \iff \delta A[x(t)] = 0, \quad A[x(t)] = \int_{x(t)} F(x(t)) \dd{x_t}.
\end{align*}
See \cite{classical-physics, path-integrals} for more details. There is no need, here, for assumptions as to steady state densities or stationary modes: this is a straightforward path of least action, where autonomous states minimise their action.

A central application of this formalism is \textit{active inference}, where the path of active states is a minimiser of \textit{expected free energy} \cite{barp_geometric_2022}.\footnote{It is important to note that one can take a path of stationary action over a system that is not stationary, by moving to paths as opposed to states.} Denoting the action of a conditional probability density as $A[-|-]$, we can formulate active states as minimisers of an action,
\[
\dot a(t) = f_a(s, a) \iff \delta A[a(t) \mid s(t)] = 0,
\]
and thus of the expected free energy (EFE)
\[
G[a(t)] = A[a(t) \mid s(t)]
\]
under certain exchangeability conditions \cite{friston2022free, path-integrals}. Note that EFE is very different from the variational free energy explored in the density-over-states formulation of the FEP, in that EFE is not a bound on surprisal. This is consistent with the fact that the density-over-paths formulation describes systems very differently from the density over states formulation.

In short, under the FEP, we can say in full generality (that is, without making any assumptions about stationarity or steady state) that the existence of a particular partition, defined in terms of sparse coupling of paths of internal and external subsets of a particular system, licences an interpretation of the most likely internal and active paths, given sensory paths, as instantiating an elemental form of Bayesian inference.\footnote{It has been claimed that the FEP \textit{requires} assumptions of stationarity and/or steady-state, or related assumptions such as ergodicity; see, e.g., \cite{di2022laying, colombo2021non}. Based on this claim, \cite{di2022laying} have suggested that the FEP cannot be used to model \textit{path-dependent dynamics}. This is problematic on two counts. First, as we have seen, the FEP in its most general form does not require that we make these assumptions. Second, when we do make the (admittedly more restrictive) assumption that the system has a NESS density (i.e., that its mechanics have a steady state solution), we can leverage the Helmholtz decomposition (examined in the next section) to split the deterministic part of the flow (or drift) into a path-dependent, conservative flow and a path-independent, dissipative flow; see \cite{Friston2019}.} Thus, the dynamics of such systems appear to engage in \textit{path-tracking} dynamics: autonomous paths look as if they track (i.e., predict) external paths. This is sometimes called self-evidencing \cite{RN92, Sakthivadivel2022b}. This has both a sentient aspect (i.e., responsive to sensory information; \cite{Ramstead2019}) and an elective or enactive aspect (i.e., decision- and planning-related; \cite{RN1096}). Formally, these are attributed to the internal and active paths, respectively; and in the case of the latter, following a path of stationary action (i.e., minimising expected free energy) is known as \textit{active inference}. 

There are two aspects to the notion of prediction in this setting. The first is that these equations of motion constitute mechanical theories, as we have defined them; they provide the laws of motion that explain the dynamics of a system. As such, they constitute a predictive (i.e., generative) model that allows \textit{us} (as experimenters or modellers) to predict the behaviour of systems, given some initial conditions, as in \cite{isomura2022experimental}. A complementary aspect is the kind prediction in which the particles that we are modelling engage. Briefly, if a system has a particular partition, then it will look as if subsets of that system (i.e., internal and external states) track each other, or equivalently, infer the statistical structure of each other. This allows us to make predictions (as modellers) about the kinds of inferences or predictions that are made by particles themselves.

We should note that this result is more minimalistic than the ones we review next. In particular, it says nothing about the specific form of the flow taken by autonomous states (indeed, this formulation leaves open an entire equivalence class of trajectories that minimize the variational free energy equally well, a more general problem for inference \cite{dill}.) 

\subsection{From paths of stationary action to density dynamics: Applications of the free energy principle to systems with a steady state solution}

This section moves to the density dynamics formulation of the FEP. Recent literature on the FEP (circa 2012-2019) has tended to focus on the density dynamics formulation, which defines a \textit{probability density over states} that evolves over time, as opposed to a probability density over paths. In this setting, we are still dealing with paths of stationary action. However, in the new setting, we assume that the statistics of the underlying probability density have a something called a \textit{steady state solution}. We discuss a result known as the \textit{approximate Bayesian inference lemma} (ABIL), which takes on two main forms in the literature, depending of the statistics of the target system. Additional assumptions can be made about the about sparse coupling and conditional independence of a particular system. When they are, they are helpful in that they make some of the mathematical derivations simpler and also, importantly, in that they allow us to say \textit{more informative things about the flow} of target systems. 

The density dynamics formulation has been explored for a number of reasons. The first is that non-equilibrium steady state, which features the breaking of detailed balance and a recurrence implied by solenoidal flow (see below), is an interesting model of biorhythms and other biological regularities. As such, looking at how far one can go under the assumption of a steady state density is an interesting exercise for modelling purposes. Since the path-based formulation is asymptotically equivalent to the density dynamics formulation (i.e., since maximum calibre is asymptotically equivalent to maximum entropy), nothing is lost by looking at the more restricted, special case\textemdash provided, of course, that the limitations of this approximation are noted. The second reason is pedagogical: it is that the derivations of the relevant equations of motion are less involved in ordinary coordinates when one assumes a steady state density.

The reader should note that, perhaps a bit confusingly, to say that a system is at steady state, or has a steady state solution, is \textit{not} a characterisation of the states of a system per se, but rather a characterisation of the \textit{time evolution} of the \textit{underlying probability density} of the system. To say that a system has a \textit{steady state solution} means that, if left unperturbed, the system would flow on the manifold defined by that solution until it arrives at a stationary point\textemdash or orbit\textemdash of that solution, where the variation of the action is zero. To say that the system \textit{is at} steady state, in turn, means that the density dynamics have stopped evolving and is now \textit{at}\textemdash (or \textit{near}, in the presence of random fluctuations)\textemdash a stationary point of its action functional, at which the action cannot further be minimised. It does not mean that the system has evolved to a fixed point. 

More formally, when the FEP is applied to the density over states, as opposed to paths, we assume that the equations of motion for the system admit a \textit{non-equilibrium steady-state density} (NESS) density. Formally, a NESS density is a stationary solution to the Fokker-Planck equation for the density dynamics describing \eqref{eq:stoch_blanket}. The assumption that this density exists makes the FEP less generally applicable (how much less is still being debated, see \cite{miguel} and the responses to that paper); but under these conditions, it can be used to say interesting things about \textit{flow} of self-organising systems. 

Under the FEP, a NESS density satisfies the following properties:

\begin{enumerate}
    \item The NESS density obeys the dependency structure of a Markov blanket \cite{friston2022free, heins2022sparse}, i.e., 
    \begin{align} \label{eq:dependencies}
        p(\mu,\eta,b) &= p(\mu | b) p(\eta | b) p(b).
    \end{align}
    \item The flow vector field $f$ under the NESS density can be written, via the Helmholtz decomposition (see Figure 2), in the form \cite{Friston2021a,DaCosta2021}
    \begin{equation}\label{decomp-eq}
    (Q(x) - \Gamma(x)) \grad J(x) - \Lambda(x)
    \end{equation}
    where $Q(x)$ is a skew-symmetric matrix, $Q(x) = -Q(x)^\textsf{T}$, $\Gamma(x)$ is positive semidefinite matrix, $J(x)$ is a potential function and $\grad$ denotes the gradient operator. $\Lambda(x)$ is a term of the form 
    \begin{align*}
        \Lambda(x)_i &= \sum_{j=1}^d \frac{\partial}{\partial x_j} \left( Q_{ij}(x) - \Gamma_{ij}(x) \right)
    \end{align*}
    where $i$ is used to index the variables in $x$. $\Lambda(x)$ contains the sum of the partial derivatives of each entry in both $Q(x)$ and $\Gamma(x)$. It has been introduced as the ``housekeeping" or correction term in the FEP literature \cite{Friston2021a, friston2022free}. 
    \item The potential function $J(x)$ in \eqref{decomp-eq} equals the surprisal; that is, $J(x) = -\log p(x)$, where $p(x)$ is the NESS density satisfying \eqref{eq:dependencies} \cite{Friston2021a,DaCosta2021}.
\end{enumerate}

\begin{figure}[t]
    \centering
    \includegraphics[width=1.0\textwidth]{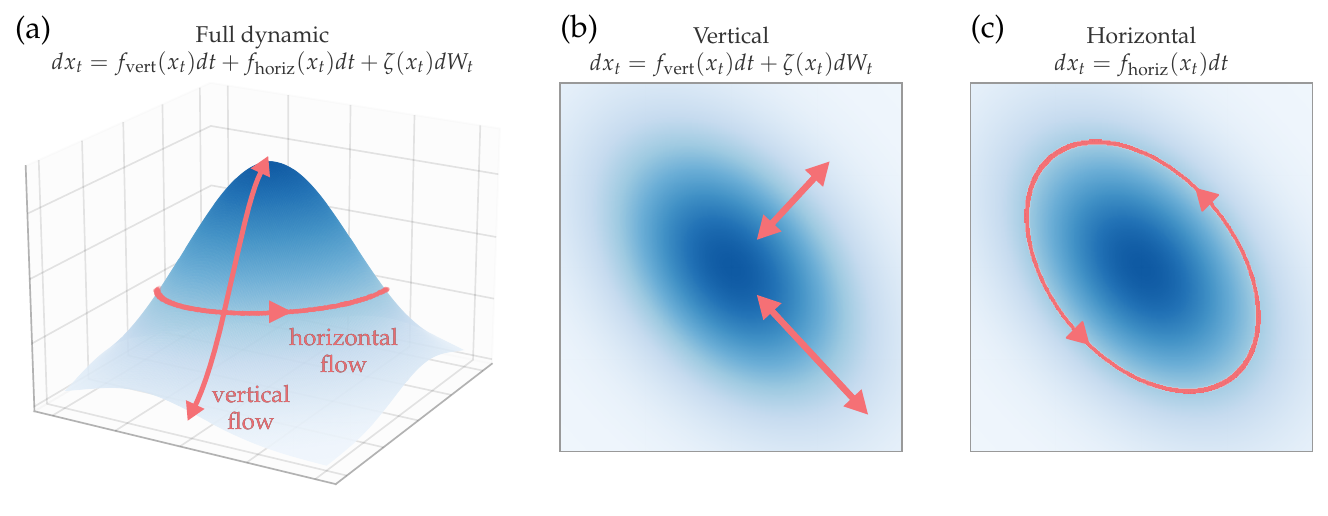}
    \caption{\textbf{Helmholtz decomposition}. Splitting of flows referred to as the Helmholtz decomposition. The vertical direction consists of a gradient ascent given by $\Gamma$ and random fluctuations pushing the system away from a mode (preventing the system from collapsing to a point). The horizontal flow is a solenoidal, energetically-conservative but temporally-directed flow, given by a matrix operator $Q$.}
    \label{fig:horiz_vert}
\end{figure}

The NESS assumption is interesting in particular systems because it allows us to say something very fundamental and informative about the kinds of flows that one finds in such systems. In these formulations, the NESS density functions as a \textit{potential function} of a Lagrangian for the system's dynamics \cite{friston2022free}. 

Crucially under the FEP, the surprisal, defined in the third point of the above definition of the NESS density, can be cast as the \textit{ontological potential} of a system (when it exists). We define an ontological potential as an abstract potential that induces an attractor for the dynamics of some system. It is ontological in the sense that it characterises \textit{what it is to be} the kind of system being considered. This is simply because the system is attracted to sets of states or paths that are characteristic the kind of system that it is, by definition (since they are attractor regions of that system). 

An ontological potential can also be written as a set of constraints on what constitutes a system-like state. Indeed, as we will see later, for the maximum entropy solution to an inference problem, the log-probability is equal to the constraints on the particular system\textemdash so it is also a potential in the literal sense of constraining the particular system to visit a set of characteristic states. That is, mathematically, we can think of the surprisal as a potential, analogous to a gravitational or electromagnetic potential, the gradients of which allow us to specify the forces to which the particular system is subject. These determine its evolution in state space, as well as in the conjugate belief space. (Correspondingly, the log-probability of a probability density constrained to weight states in a certain way reproduces that weight, i.e., that 
\[
p(x) = \exp{-J(x)} \iff \log p(x) = -J(x).
\]
We will explore this further in the notion of an ontological \textit{constraint}, which is dual to the notion of an ontological potential, and introduced in Section \ref{sec:maxent}.)

When we consider the statistics of sampling dynamics that converge towards the \textit{mode} of an intended NESS density, the ontological potential acquires another interpretation in terms of the system's \emph{preferences} \cite{Friston2019,efe,millidge2020relationship,attias2003planning,rawlik_stochastic_2012,levine_reinforcement_2018}. We can view the NESS density as providing a set of \textit{prior preferences} which the particular system looks as if it attempts to enact or bring about through action \cite{DaCosta2021}. Indeed, we can think of this solution to the dynamics as a naturalised account of the teleology of cognitive systems \cite{levin-TAME, RN1250}. 

With these assumptions in play, we can derive a stronger version of the claim that particular systems engage in a form of approximate Bayesian inference. This approximate Bayesian inference lemma (ABIL) can be stated as follows: when a system has a steady state solution, we can define a \textit{synchronisation map} that systematically relates the \textit{conditional mode} of external states to that of internal states.\footnote{Note that this map induces what is termed a {\it synchronisation manifold}, which is where the synchronised trajectories are collected, see e.g. \cite{synch}.} Under these conditions, we can say that the particular looks as if it performs inference about an optimal conditional mode, by internally encoding the statistics of the outside environment. The ABIL itself says that under a synchronisation map and a variational free energy functional (or its equivalent), this mode matching is both necessary and sufficient for approximate Bayesian inference \cite{Sakthivadivel2022b}.

We can define the synchronisation map $\sigma$ formally. The map $\sigma$ is a function that sends the most likely internal state, given a blanket state, to the most likely external state, given that same blanket state. These internal states are what really flow on variational free energy, a variational ``move'' that allows us to talk about \textit{inference}\textemdash since this flow shares the same minimum as the flow on surprisal, we can read these states as performing inference. As we said, this licenses an interpretation of the dynamics of a system as instantiating an elemental form of approximate Bayesian inference \cite{Parr2022a}.

Let $\hat\mu_b$ be the \emph{maximum a posteriori} estimate $\arg \max p(\mu \mid b)$, and $\bm\mu(b)$ be a function taking $b$ to $\hat\mu_b$ (and identically for $\eta$). Usually, there exists a synchronisation map 
\begin{gather*}
\sigma(\hat\mu_b) = \hat\eta_b\\
\hat\mu_b = \arg \max p(\mu \mid b)\\
\hat\eta_b = \arg \max p(\eta \mid b)
\end{gather*}
described in \cite{Friston2019, miguel, parr_markov_2020, DaCosta2021}. We can depict this relation in the following diagram (adapted from \cite{DaCosta2021}):
\begin{equation*}\label{fig:comm-diag}
\begin{tikzcd}[row sep=6em,column sep=7em]
 b\in\mathcal{B} \arrow[d, swap, bend right=20, "\bm\mu"] \arrow[r, "\bm\eta"] &\hat\eta_b  \\
\hat\mu_b \arrow[ur, swap, "\sigma = \bm\eta\circ\bm\mu^{-1}"] \arrow[u, swap, bend right=20, "\bm\mu^{-1}"]
\end{tikzcd}
\end{equation*}
where $\mathcal{B}$ is the set of possible blanket states, and where we have assumed, for illustration's sake only, that $\bm\mu$ is invertible.

To elaborate more informally, what this means is that, for every blanket state, there is an average internal state or internal mode that parameterises a probability density over or belief about an average external state or external mode \cite{DaCosta2021}. The claim behind the ABIL is that systems which match these conditional modes across a Markov blanket are storing models of their environment (or, can be read as such) and thus are engaging in a sort of inference. One should note that the existence of $\bm \mu^{-1}$ is not guaranteed: we can prove that $\sigma$ exists if and only if $\bm \mu^{-1}$ is invertible {\it on its image}, i.e., is an injection (see \cite{DaCosta2021} or \cite{Sakthivadivel2022b}), but {\it a priori} cannot claim that $\bm \mu^{-1}$ necessarily exists on any domain. That said, it is important to also note that we do not at all need $\bm \mu$ to be bijective.

Furthermore, note that\textemdash given its dependence on the existence of a NESS solution to the mechanics of a particular system\textemdash this description is only taken to hold in the asymptotic limit \cite{ying-jen, touchette, paths-comment}. 

Now, the conditional external mode may or may not have any interesting dynamics to it. This is where we encounter second arm of the typology of formal applications (see Figure \ref{fig:schematic_three_branches}), as the FEP applies informatively to the former case, and vacuously to the latter. For instance, in the class of one-dimensional linear systems analysed by \cite{miguel}, there are no dynamics to the mode at all: the only source of variation in the flow is the random fluctuations. As discussed in \cite{paths-comment}, in linear systems, the dynamics simply dissipate to a stationary \textit{fixed point} and remain there. Now, a system that conforms to the FEP will still match the external mode via the synchronisation map, but since there are no dynamics to the external mode, by construction, there can be no proactive ``tracking'' that might be interpreted as sentient, proactive sampling. Accordingly, \cite{miguel} find that free energy gradients in these cases are uninformative about the real dynamics of the system; but this is because \textit{there are no dynamics} about which to say anything interesting. We can think of this behaviour as \textit{mode matching}, a kind of static Bayesian inference (e.g., apt to describe Bayesian inference that statisticians apply to data under a general linear model).

In systems where there \textit{is} a dynamical aspect to the external mode, we instead obtain a much richer, proactive kind of \textit{mode-tracking} behaviour, where the external state is changing over time. In these conditions, the internal mode will seem to be \textit{tracking} the external mode. Since mode tracking entails the most likely flow following the deterministic component of equation \eqref{eq:stoch_blanket}, this is like the classical limit\textemdash i.e., the limit of infinite certainty\textemdash of the path integral of free energy \cite{friston2022free}. That is to say, certain mode tracking particular systems are macroscopic Bayesian ``particles'' for which we can ignore random fluctuations. 

The following of a mode as beliefs change induces a conjugate \textit{information geometry} and a corresponding flow on the conjugate statistical manifolds, in the sense that such a system performs inference at each time point to determine what belief its internal state should be parameterising, and flows towards that optimal parameter \cite{parr_markov_2020, paths-comment}. As indicated, in one case, we consider the ``intrinsic geometry'' of probability densities defined over the physical states (or paths, i.e., sequences of states) of a system, i.e., the probability \textit{of} these states or paths; and in the other case, we consider the ``extrinsic geometry" of probability densities that are \textit{parameterised by} these states or paths, i.e., we treat them as the \textit{parameters of probability densities} over some other set of states or paths. See \cite{Friston2019, RN1165}.

In the next section, \ref{sec:math_prelim}, we will see that we can use the technology of the constrained maximum entropy principle (CMEP) to reformulate the ontological potential (i.e., the NESS potential) as a set of \textit{constraints} against which entropy is maximised as the system dissipates. 

\subsection{Some remarks about the state of the art}

Before turning to the CMEP and its connection to the FEP, we comment on some important developments in the technical literature on the FEP. Recent work \cite{miguel, biehl2021technical} has questioned whether Markov blankets are as ubiquitous as claimed by theorists of the FEP. We briefly address this work. In summary, we argue that Markov blankets (as defined in the appropriate sense, in terms of a particular partition) are ubiquitous in physical systems: essentially all physical systems feature Markov blankets. 

According to the so-called sparse coupling conjecture (SCC), all sufficiently large, sparsely coupled, random dynamical systems have a Markov blanket, defined in the usual way. Recent work has shown that the SCC holds generically in an approximate form for systems with quadratic surprisals (including quadratic surprisals with state-dependent Helmholtz matrices). That is, we now know that as an extremely generic class of random dynamical system increases in size (i.e., as they become higher dimensional), the probability of finding a Markov blanket in the system, defined in the appropriate way (i.e., between subsets of a particular partition), tends to one.

In \cite{weak-blankets}, a weakened version of the SCC is proven. The cited results show that the Hessian condition used to investigate Markov blankets, even in a large class of non-linear systems, is obtained as dimension increases. Those results build on previous work \cite{heins2022sparse}, which identified a sufficient condition for claiming that a system displays a Markov blanket in the case of systems with Gaussian steady-state densities. This condition is that the inner product of the Hessian of the steady state distribution of the system (the entries of which encodes the curvature, or double partial derivatives, of the surprisal) and the matrix field that captures solenoidal part of the flow be identically zero. When this inner product is identically zero, we always have a Markov blanket in the appropriate sense. Now, in \cite{heins2022sparse}, it was merely \textit{conjectured} that the probability of finding a blanket increases with the size of the system considered. The intuition was that, as a system increases in size, there is more ``room'' for it to be sparse, and thereby, to evince a Markov blanket between subsets. In \cite{weak-blankets}, it is proven that the Markov blanket property holds with probability one for many coupled random dynamical systems of sufficient size. The proof involves defining a ``blanket index,'' which scores the degree to which the inner product discussed is nonzero. Using this technology, one can explicitly quantify the degree to which systems depart from the strict Markov blanket condition. More interestingly, the probability of the blanket index vanishing tends to one with dimension. Crucially, most physical systems are large in the relevant sense. A mere teaspoon of water, for instance, contains approximately $10^{23}$ molecules. The brain contains 100 billion neurons, with each individual neuron making thousands of connections. Other examples abound.

Now, it may be true that the results in \cite{biehl2021technical} undermine the original derivation of the approximate Bayesian inference lemma (ABIL), as it can be found in the well-known paper \cite{friston2013life}. However, newer work has re-derived the ABIL using conventional mathematics \cite{Sakthivadivel2022b}. The results in \cite{biehl2021technical} only pertain to the derivations found in \cite{friston2013life}; and we note that the latter is also critically discussed in \cite{friston2021some}. Thus, the appropriate conclusion to draw from \cite{biehl2021technical} is that one ought not to cite \cite{friston2013life} to make points about the ABIL or the Markov blanket property; and more subtly, that one should move on from that formalism. But we have independent reasons to believe that the ABIL is true; and indeed, the literature has moved on from that formalism.

As discussed above, it would be misleading to conclude that the FEP does not apply to the systems analysed in \cite{miguel}. The application of the FEP is uninformative because that work focuses on linear, low dimensional mathematical edge-cases, i.e., systems with a small number of states \cite{paths-comment, karl-comment}. More precisely, the paper considers whether the FEP can be applied usefully to one-dimensional dissipative systems; physically, these are coupled, dampened springs with one degree of freedom each. The paper cogently shows that it is difficult to construct a Markov blanket for such systems. However, this does not undermine the FEP or the obtaining of the Markov blanket property in the general case. Instead, these results constitute an interesting application of the FEP to very low (i.e., one) dimensional, linear systems. As such, the conclusion to draw from this work is not that the Markov blanket property does not obtain in general, but rather, that Markov blankets are rare or difficult to construct in small, low dimensional systems; but they remain ubiquitous in appropriately large ones (which comprise most physical systems). Thus, the FEP is not a ``theory of everything'' in the sense that it could be applied informatively to kind any mathematical system whatsoever\textemdash it is only a theory of everything that features a Markov blanket. Indeed, \cite{miguel} have shown that the FEP applies vacuously to all sorts of systems; e.g., it applies to, but has nothing particularly interesting to say about, linear stochastic systems. This is by design: there is nothing interesting to say, FEP-theoretically, about such systems.

In summary, the FEP is a method or principle that applies to ``things'' that are defined stipulatively in terms of Markov blankets, sparse coupling, and particular partitions. The FEP is not concerned with systems that do not contain any ``thing,'' so defined. On this view, the critical literature above focusses on whether or not any given system can be partitioned into some ``thing'' and every ``thing'' else. If it can, then the FEP applies\textemdash but not otherwise.
\section{Some mathematical preliminaries on the maximum entropy principle, gauge theory, and dualization}\label{sec:math_prelim}

In the introduction, we discussed how changing our perspective on self-organisation entailed an exchange of points of view across a boundary: rather than asking how a particular system or particle maintains its ``self'' and what beliefs it ought to hold about the environment, as the FEP is interested in, we can instead ask what that self is, and what it looks like from the perspective of an outsider observing the system. Likewise, to dualise the objects that we ask about also implies dualising our application of Bayesian mechanics, to ask about our beliefs about a system, rather than the beliefs encoded or carried by the system. The idea, then, is that we can leverage this dual perspective to model self-organisation as we conventionally would, restoring the symmetry of the problem and allowing us to apply the FEP to model self-organised systems.

We have said that the FEP is dual to the constrained maximum entropy principle (CMEP). Duality in this category-theoretic sense means that two objects, formally called an adjoint pair, share some common set of intrinsic features\textemdash but exhibit \textit{relationships} to other objects \textit{in opposite directions}. An adjunction (the existence of an adjoint pair) usually suggests some interesting structure hiding in a problem; in this case, it is the peculiar agent-environment symmetry which is definitional of coupled systems that infer each others' states. It can be proven that exchanging (i) free energy for constrained entropy, and (ii) internal for external states, recovers all aspects of the ABIL and a simple case of self-evidencing, and thus much of the FEP, especially as it pertains to self-organisation; see \cite{Sakthivadivel2022b} for a proof of the ABIL (Lemma 4.2 and Theorem 4.1).

Thus, the motivation for dualisation is almost three-fold: (i) it recaptures the original spirit of the FEP, which is that of an observer modelling an agent exhibiting self-organisation; (ii) it allows us to ground the mathematics of the FEP in the well established foundations of maximum entropy and stationary systems at equilibrium; (iii) it allows us to extend the existing methods of the FEP to scenarios that are new to the FEP literature, such as constraint-based formalisms. As a technical tool, changing our viewpoint introduces \textit{constrained self-entropy} as the dual to the free energy of beliefs. In doing so, we can relate the FEP to existing insights in probability theory and dynamical systems theory. This new viewpoint turns out to be independently interesting for our reading of the FEP, and possibly extends it to new phenomena or systems. New approaches developed for the maximum entropy principle (such as gauge-theoretic results) reflect themselves in useful ways\textemdash within the FEP\textemdash via this relationship.

We begin not with maximum entropy, \emph{per se}, but with a somewhat unconventional \textit{geometric view} on the CMEP, to be related later to the density over states formulation of the FEP (and especially the Helmholtz decomposition). The core elements of the CMEP needed for this particular construction have their origins in \textit{gauge theory}, a theory in mathematical physics that relates the dynamics of particles to the geometry of their state spaces. Later, this will allow us to discuss the decomposition of flows under maximum entropy in the same fashion as what exists under the FEP, as well as tie it in explicitly to the updating of probability distribution in response to changes in constraints. We recommend \cite{baez} as a reference with a good introductory tone, and \cite{rubakov} or \cite{Nakahara2003} for more details.

A gauge theory begins with a field theory like electromagnetism or quantum electrodynamics (QED), describing the dynamics of matter and the particles that comprise it. The dynamics of ``matter fields'' are typically described by applying a principle of stationary action (cf. Section \ref{sec:preliminaries}) and thus are related to a special integral called an action functional, which, as we have discussed in Section \ref{sec:preliminaries}, is a quantity that gets minimised by the field (or in quantum field theory, a description of the field's most likely state). A functional, as we have said, is a function of a function: in this case, the action functional is a function of the Lagrangian of a system, which to repeat, summarizes the energies involved in the mechanics of the system. It follows that the minimiser of the action functional, which is a point in a function space, gives us the configuration of the field where the action is stationary.

In principle, the action functional gives us everything we need to know about a matter field. However, in many field theories, the action admits some sort of \textit{symmetry}\textemdash this is a transformation that leaves the action \textit{invariant}, so that an \textit{arbitrary change} in some particular quantity has no effect on the equations of motion of the matter field predicted by the action. A loose example includes the gravitational field under changes in reference frame: the principle of general relativity is that we do not have absolute coordinates in which to describe physics, and things like motion appear different from different perspectives, despite the underlying physics being the same. Hence, gravity has a coordinate invariance, which means it stays the same under changes in coordinates: in other words, it has a symmetry under this transformation. In other theories, we have other symmetries: for instance, in QED, we can choose and change the complex phase of a particle arbitrarily, with no change in the associated action. The quantity under which the theory is symmetric is called a \textit{gauge}. In gauge theory, the symmetry itself is referred to as \textit{gauge invariance}, which is characterised by a free choice of gauge and invariance under changes in gauge called gauge transformations. 

\begin{figure}[H]
    \centering\includegraphics[width=1.0\textwidth]{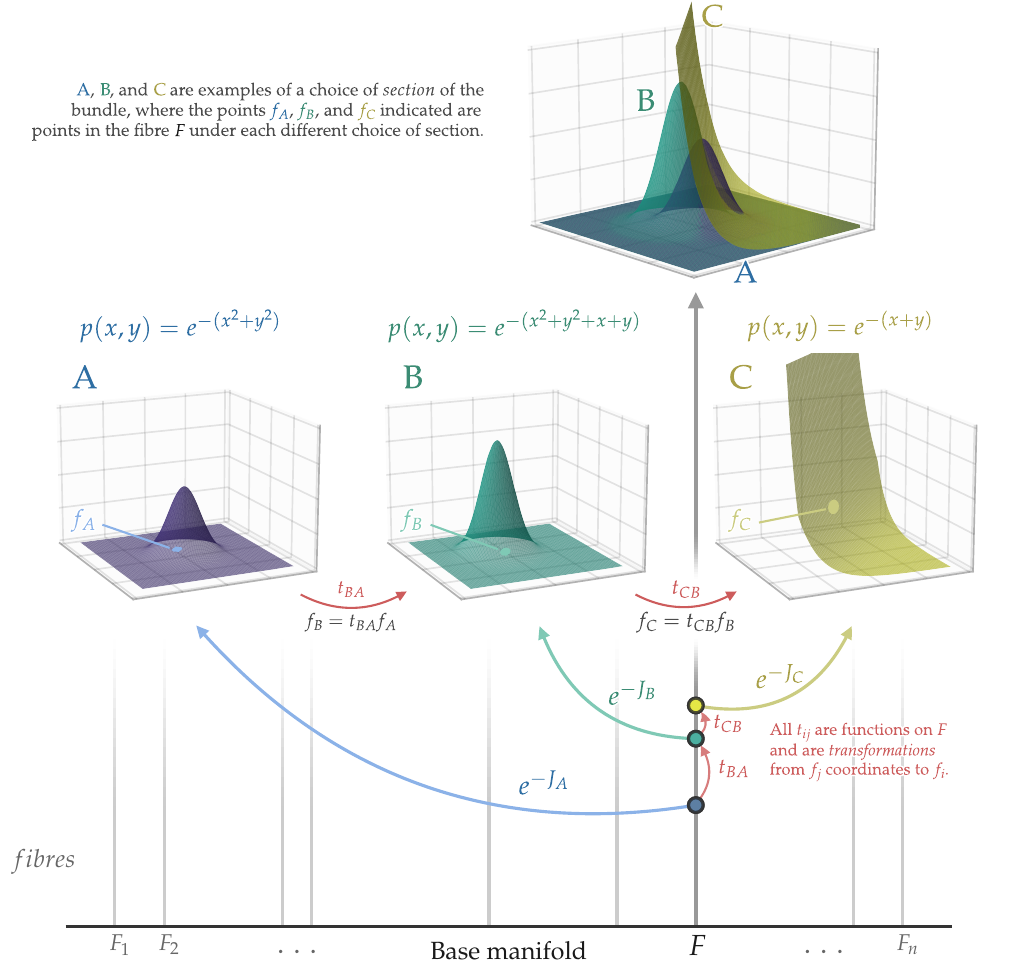}
    \caption{\textbf{Illustration of fibre bundles, sections, and transformations}. A fibre bundle exists on a base manifold, situated over points in that space. Here, the fibre is a copy of the real line attached to each point in the base. Different choices of \textit{section} of the bundle correspond to different choices of constraint function. The section is a function which assigns coordinates in ``probability space'' --- by which we mean values of probability, or specific points in the fibre $F$ over some base point $(x,y)$ --- to points on the base. For $f_A, f_B, f_C \in F$, these points lie over the base point $(x,y)$. The height of any $f_i(x,y)$ corresponds to the probability of $(x,y)$, and is a lift of $(x,y)$ within the section $p$. Three simple choices of section are shown, with the same point $(x,y)$ mapping to different regions of the corresponding probability density due to a different choice of constraint function ($A$, $B$, and $C$). The inset over $F$ shows that all of these densities are sections over the same base space.}
    \label{fig:covariance_fig}
\end{figure}

One of the reasons why gauge theories are interesting is because, unlike the action functional, the matter field itself is generally gauge \textit{covariant}; this means that it changes with a change in gauge. Whilst we can, in principle, deduce all the information we need about a matter field from the action functional, this symmetry is not manifest in the field: the equation expressing the evolution of the field changes with the choice of gauge. This is what is meant by covariance (varying together). Imagine choosing a reference frame: gauge symmetry only says that we can choose any new frame and still observe motion consistent with the laws of physics (e.g., motion which conserves total energy); but, the expression of that motion within a frame is still dependent on the choice of frame (for instance, choosing a moving frame of reference will convert inertial trajectories into moving ones, relative to that frame of reference). Gauge covariance has a very literal relationship with the idea of changing the coordinate basis in which we express the components of a vector. (For an example of this see Figure \ref{fig:covariance_fig} on the following page.)

The tension between physical arbitrariness and mathematical relevance is what is captured by gauge theory, and correspondingly, gauge theory gives us a way of speaking about how one quantity varies another as it moves on space-time. In gauge theory, this often records the way that a force field varies the motion of a particle, and it records the coupling of bosons (force-carrying particles) to fermions (particles that comprise matter). An instructive example is that of gravity, under general relativity. It is the principle of \emph{general} relativity that special relativity (i.e., the relativity of motion with respect to the speed of light) can be extended to any expression of motion which is accelerating relative to some other expression of that motion. A consequence of this view is that all non-inertial motion is identical to inertial motion on a curved surface which causes some acceleration, called the ``equivalence principle.'' The equivalence principle was properly introduced by Albert Einstein in 1907, when he observed that the acceleration of bodies towards the center of the Earth at a rate of $1g$ is equivalent to the acceleration of an inertially moving body (i.e., not accelerating \emph{within} a frame of reference) that would be observed on a rocket in free space being accelerated at a rate of $1g$ (its \emph{frame of reference} is accelerating). Again, this is the equivalence principle: accelerating reference systems are physically equivalent to gravitational fields. As such, the curvature of space-time is gravity. As summarised in \cite{giancoli}, ``there is no experiment observers can perform to distinguish whether an acceleration arises because of a gravitational force or because their reference frame is accelerating." Now, it is mass that deforms space-time, completing the analogy. Another truism, due to John Wheeler, is that ``space-time tells matter how to move; matter tells space-time how to curve.''

The way mathematicians speak about gauge theory is via a special kind of space called a \textit{fibre bundle}. There are three ingredients to our gauge theory: an ``associated bundle" $E$ with fibres $F$, where the matter field lives; the space that the matter field lies over $X$, and the choice of gauge for the matter field, which lives in a ``principal" bundle $P$. The choice of gauge has the specific property that it transforms the matter field when it transforms, so these two degrees of freedom are linked, and both lie over some input or base space, so those are linked too. 

Each field has a field state at a point over the input space, so we think of this three-fold structure in terms of a space containing all of the possible states of the matter field attached to each point in the input space. This is a fibre bundle, so named because it looks like a set of fibres sticking out of a base manifold, bundled all together. We do the same for the choice of gauge at every input point, and then couple the two bundles by associating matter field states with choices of gauge.

At its mathematical core, the fibre bundle construction is straightforwardly like a generalisation of a \textit{function}: at each input, there is a whole space of possible outputs or images, which we bundle together across the whole input space. An example of this is the fact that the $xy$-plane is a bundle of real lines over points on the real line, arranged topologically such that each fibre is at a right angle to the base, which allows us to define real-valued functions of a single real input. More complicated functions have a natural home in this framework by changing what our fibres and input space are. This includes the states of classical and quantum fields on space-time. See Figure \ref{fig:constraint_geom}.

We refer to the internal functions in a bundle, such as $f(x)=y$ in a real line bundle over the real numbers $\R$, as \textit{sections}. One can imagine the image of a section, for instance, a set of $y$ values tracing out a path in the plane parameterised by points in the base, as a cross-section of the bundle, slicing it along that path. Thus, sections are functions that generate slices of the bundle. Sections lift paths in the base to paths in the bundle to produce such slices. Indeed, what we have called a slice of the bundle (the image set of $f(x)$, consisting of a set of a particular set of $y$ values) is actually called a \textit{lift}.

Bundles generalise functions between spaces, and in particular, a bundle allows one to construct a function from a base space to a bundle of other spaces called fibres. This allows us to define \textit{fields as sections of a fibre bundle}, since they are functions that `reach' into a bundle and pick a field state at an input point on the base space. For example, a classical field is a section of some bundle: on every point of space-time, we get a classical state, such that a classical field is a lift from space-time into a field of states. This example falls along the same lines as our distinction between mechanics and dynamics, i.e., restricting a field to lie on a certain line in space-time produces mechanics, and feeding the section on that line the input points we desire under some form of the section gives a trajectory (i.e., dynamics). 

Taking a path on space-time and lifting it into, say, a complex line bundle gives us the complex phase of a quantum particle. What that particle \textit{does} is determined by the actual equation of motion for the lift (its mechanics), which is the degree of freedom that we identify with specifying a mechanical theory for whatever particle it is. In turn, the mechanical theory itself arises from the \textit{existence} of such a lift. The full picture, then, is that we have a fibre bundle over space-time, which gives us a field-theoretic structure, whose lift at a point is a mechanical theory. Indeed, a section of a complex line bundle is a wavefunction, and inputting data like a particular potential function gives us a quantum equation of motion.

The fact that the expression of the function\textemdash the frame of reference for the matter field\textemdash is linked to the choice of gauge by construction is the \textit{gauge covariance} we defined previously. Refer back to the three-fold structure we defined earlier. 

The last ingredients to complete our presentation of gauge theory are the \textit{gauge field}, \textit{gauge force}, and \textit{connection}. A connection is a generalisation of the derivative that allows us to talk about how a choice of section varies across the base manifold. Defining a connection is like inducing a fine topology on the total space of the bundle, allowing us to map infinitesimal changes in the base space to infinitesimal changes in the bundle space. Thus, we can take derivatives of paths through the bundle which are parameterised by a path in the base. In this setting, the derivative itself is a generalised object called a tangent space, which is a collection of tangent vectors (a vector field) which describes the ways that a particle can flow from a given point in the base space. In gauge-theoretic terms, the connection is the gauge field, which tells us how a choice of gauge varies across space-time. Paths which are flat in the connection are unforced, whilst paths that curve out or are deflected from a flat plane, and are experienced by the particle as a \textit{gauge force}.

The connection also allows us to vary the constraints on motion in the bundle space, and, allows us to define \emph{parallel transport} of points therein. We will discuss parallel transport in particular later. We can now use this to make sense of the covariance of $p(x)$ and $J(x)$ in maximum entropy, providing a justification grounded in physics for the \textit{unreasonable effectiveness of approximate Bayesian inference}, which we turn to next.
\section{On the duality of the free energy principle and the constrained maximum entropy principle}\label{sec:maxent}

In this section, we leverage gauge-theoretic resources presented in Section \ref{sec:math_prelim} to provide a complementary (i.e., dual) perspective on Bayesian mechanics under the FEP. We provide a brief recapitulation\footnote{For clarity, note that this section is meant to provide a review of existing work, rather than presenting new results.} of results from \cite{Sakthivadivel2022a, Sakthivadivel2022b}, following \cite[chapter 9]{Nakahara2003} as a mathematical reference for gauge theory. This section has two parts: moving from constraints to gauge symmetries, and from gauge symmetries to dynamics.

In this dual version of Bayesian mechanics, the ontological potential of a system is expressed as a set of constraints on the states of a system that one can specify using a variational \textit{maximum entropy} procedure. This contrasts with the NESS potential over states, which is written in terms of sensory causes for those states. Correspondingly, an ontological potential specifies the states that the system is likely to find itself in (i.e., states that are typical of the system). However, this does not tell us much about the systems dynamics towards those states; these are largely governed by the system's horizontal (typically called solenoidal) and vertical (dissipative) flows (see Figure \ref{fig:horiz_vert}). We will discuss that here.

\subsection{From constraints to gauge symmetry}

\begin{figure}[htb!]
    \centering
    \includegraphics[width=0.8\textwidth]{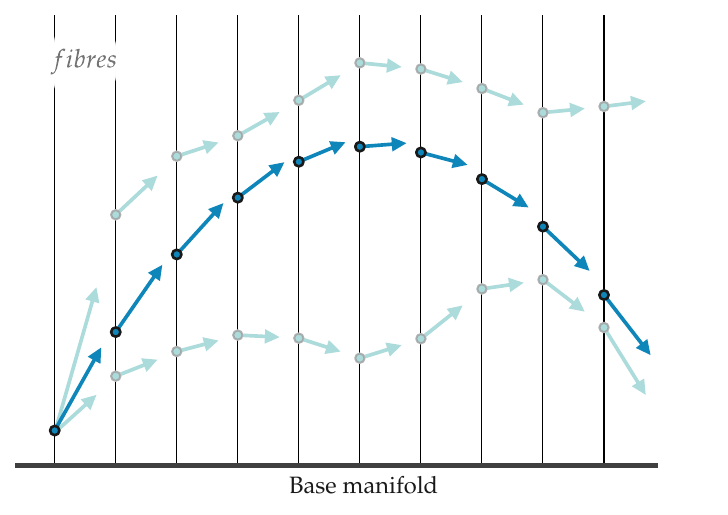}
    \caption{\textbf{Vector field on a curve.} A connection in a bundle allows us to define a tangent vector field along a curve. Here, the constraints applied are directly constraints on the shape of such a curve, applied as a vector field constraining the motion of some particle to lie in the gradient of a potential. This particular path is curved, corresponding to a least action path\textemdash a geodesic with respect to some (curved) connection. In this local patch, we can also easily construct a flat connection with flat paths. We refer to paths that are flat as horizontal lifts, and parallel transport of points happens along these paths in particular.}
    \label{fig:fibre_tangent}
\end{figure}

We previously mentioned that, in a certain mathematical sense, gauge theory is the ideal way to speak about how one quantity varies another quantity when it changes. This covariance is precisely what occurs when probability is transported across the state space\textemdash when the constraint on a state changes, so does the probability of that state, in the same precise way as a connection constrains the motion of dynamics in a space. Consider the Lagrangian optimisation condition that the gradient of some function of interest equals the gradient of the constraints up to a scaling (a constant called a Lagrange multiplier), i.e., that
\[
-\pdv{x} \log p(x) = \lambda \pdv{x} J(x)
\]
when entropy is maximised for a Lagrangian $-\log p(x) - \lambda J(x)$. This is also known as the Euler-Lagrange equation for the maximisation of entropy, and the $p(x)$ for which this equation is true is the maximum entropy distribution. The equation demands that the evolution of the surprisal function $\log p(x)$ as a vector field is equivalent to the gradient of some other function. The aim of this section is ultimately to see how precise the analogy between the above optimisation relationship, and a potential function that constrains the movement of a particle described by the image of some section, is. We will use this as motivation to prove that constrained maximum entropy actually constrains the probability density of interest, i.e., that under fixed constraints, the density is constrained to have a shape such that the gradient on $p$ lies in the induced connection on an associated bundle. See Figure \ref{fig:fibre_tangent}.

In some sense, the importance of this is not that this is a gauge symmetry, in the conventional sense of a physical field theory (though it is, as we will see, from the perspective of the entropy functional). The power instead lies in the \textit{geometric relationships} brought to bear, and in particular, the idea that when one changes the constraints of a system, one also changes the vector field on a curve or surface telling us how our assignment of probability should ``move'' over the state space. This is in the \textit{very same} sense as of ``movement'' as is instantiated by Bayesian updating: changing our prior knowledge or constraints redistributes probability. Indeed, this is the power of the formalism: to arrive at a geometric view of Bayesian inference, both approximate and exact. 

We will first show that this gauge symmetry exists. Then, in the following subsection, we will use it to interpret gauge covariance in the context of parallel transport. Parallel transport licenses new interpretations of both belief updating and the Helmholtz decomposition (or flow splitting) in the FEP (see Figure \ref{fig:horiz_vert}), connecting it to well established foundations of mathematical physics. 

The standard entropy functional to be maximised is
\[
-\int p(x) \log p(x) \dd{x}
\]
subject to a constraint that some function on those states is on average $C$, i.e., 
\begin{equation}\label{constraint-eq}
\int J(x) p(x) \dd{x} = C,
\end{equation}
for a total action of
\begin{equation}\label{max-ent-action}
S[x;J] = -\int p(x) \log p(x) \dd{x} - \lambda\left( \int J(x) p(x) \dd{x} - C\right).
\end{equation}
By \eqref{constraint-eq}, it is suggestive\textemdash if potentially na\"ive\textemdash that the lattermost term is zero. This observation means that any given choice of $J$ factorises away. However, to produce a true gauge theory, we also have to investigate what changing a choice of $J$ does to $p$. Under a change from $J$ to some new choice of constraint, $J + J'$, we deduce the following \emph{transformation law} for \eqref{max-ent-action}:
\begin{equation}\label{trans-max-ent-action}
-\int e^{-\lambda J'(x)}p(x) \log e^{-\lambda J'(x)}p(x) \dd{x} - \lambda\left( \int (J + J') e^{-\lambda J'}p(x) \dd{x} \right).
\end{equation}
In Section \ref{sec:preliminaries}, we introduced the idea that the variation of the action gives us \emph{the} trajectory of a system, or some equation for that trajectory that can be solved like Newton's law. Initially, it follows that the action should give us a \textit{unique} motion or equation of motion for a system, and that different actions will give us different least action paths. In a gauge theory, however, the gauge symmetry expresses itself as a \textit{redundancy in the possible trajectories} of the system: there are multiple possible \emph{gauge equivalent} paths or field configurations for the system. We can show this easily if we note the Euler-Lagrange equation for
\[
\argmax_{p(x)} S[x;J]
\]
is
\begin{align}\label{max-ent-root}
\pdv{p(x)} \big( - p(x) \log p(x) - \lambda J(x) p(x) \big) &= 0 \nonumber \\
-\log p(x) - \lambda J(x) &= 0,
\end{align}
which arises simply by taking the integrand of \eqref{max-ent-action} and setting the gradient of it equal to zero. The solution\footnote{Note in the above that we have absorbed the $-1$ resulting from evaluating $- \pdv{p(x)} \log p(x)$ into the Lagrange multiplier $\lambda$. The normalisation constant $Z$ has also been absorbed into $\lambda$, as a matter of convention \cite{dill}. Hence, all $p$'s indicated here are probabilities.} to this problem is $\exp{-\lambda J(x)}$, the root of a particularly simple algebraic equation. Now we will try to do the same for \eqref{trans-max-ent-action}. Taking the same variation of $S[x; J+J']$ yields
\[
\pdv{p(x)} \big( - e^{-\lambda J'(x)}p(x) \log e^{-\lambda J'(x)}p(x) - \lambda (J(x) + J'(x)) e^{-J(x)} p(x) \big) = 0.
\]
Note that the integrand is more complicated, due to the transformation law we defined. Using properties of the logarithm, we can simplify the first term to
\[
\pdv{p(x)} \big( - e^{-\lambda J'(x)}p(x) (-\lambda J'(x) + \log p(x)) - \lambda (J(x) + J'(x)) e^{-J(x)} p(x) \big) = 0.
\]
This allows us to factorise out $e^{-\lambda J(x)}$, yielding 
\[
e^{-\lambda J'(x)} \pdv{p(x)} \big( - p(x) (-\lambda J'(x) + \log p(x)) - \lambda (J(x) + J'(x))p(x) \big) = 0.
\]
Since an exponential function is always greater than zero, we can drop this constant entirely\textemdash it has no effect on the zero point of the gradient term. Some further algebra gives us the following grouping of terms:
\[
\pdv{p(x)} \big( (\lambda J'(x) - \log p(x))p(x) + (-\lambda J(x) -\lambda J'(x))p(x) \big) = 0.
\]
We now take the derivative indicated, yielding the equation
\[
(\lambda J'(x) - \log p(x)) + (-\lambda J(x) -\lambda J'(x)) = 0.
\]
At this point in the calculation, it is obvious that the new constraint function cancels out. As such, we get our original solution,
\begin{equation*}
- \log p(x) -\lambda J(x) = 0,
\end{equation*}
back. This recovers \eqref{max-ent-root}, completing the result.

In summary, we have shown that fixing a particular constraint $J$ is \textit{arbitrary} and also that \textit{changing} that choice of constraint is \textit{also arbitrary}. Thus, the choice of constraint is a degree of freedom in the specification of the mechanics of the system that does not affect the action (i.e., changes leave the action stationary). This generalises a symmetry previously noted by Jaynes, in that re-parameterisations of a constraint should not affect the resulting probability density \cite{Jaynes1957}, which was re-introduced by Shore and Johnson \cite{shore} in their axiom scheme for consistent inferences under maximum entropy (see \cite{dill} for a review).

In fact, this symmetry is rooted in Jaynes' original claims about maximising entropy. The fact that there exists an innumerably large class of systems which are all described entropically, with particularities that can be fixed to produce a description of a specific system, gives these constraints the status of a \textit{gauge symmetry}. Just like there are certain privileged choices of gauge based on the computational features they have (e.g., the Coulomb, Lorenz, and Fock-Schwinger gauges in electromagnetism), we have specified a particular privileged choice of gauge that we may call the Bayesian gauge,\footnote{The authors thank James Glazebrook for suggesting this name.} producing a system that engages in approximate Bayesian inference. Moreover, there is exactly a suggestion of the arbitrariness of the constraints in the Bayesian gauge. Precisely, this free choice is a \textit{free choice of gauge}, for inference written down as a gauge theory. Suppose, for instance, that the sufficient statistics of a target probability density $p$ are only a mean $\hat x$. The constraint 
\[
\int q(x) \log p(x) \dd{x} = 0
\]
and 
\[
\int x q(x) \dd{x} = \hat x
\]
both result in $q = p$, the former by direct solution; the latter by recognising that amongst Gibbs measures $p= \exp{-\lambda x}$ if $\hat x$ is the unique sufficient statistic for $p$ (note we are in the world of exponential families, which is explicitly our area of interest given the construction in \cite{Sakthivadivel2022a}). 

In some sense, this gauge-theoretic relationship \emph{is} the reason for approximate Bayesian inference; i.e., the reason \textit{why} it works at all. It amounts to a statement that it is sufficient to learn the statistics of an outside world for that world to no longer be surprising. More generally, it provides a \textit{definition} of variational Bayesian inference in terms of constraints on the value of a control parameter.\footnote{Here, a control parameter is a variable (possibly exogenous to the system in question) which plays a critical role in setting the dynamics of that system\textemdash for instance, blanket states from the perspective of internal (resp. external) states.} In the language of mathematical approaches to gauge theory, we can cast this as a Bayesian gauge group, consisting of exponential functions (beliefs, probabilistically speaking) where changes in those beliefs are automorphisms of an appropriate (principal) bundle (see Proposition 1 and Theorem 2 of \cite{Sakthivadivel2022a}). We thus have a kind of mechanical theory, written in gauge theoretic terms, for all systems that look as if they engage in inference.

We can also talk about priors probabilities using this setup. The fact that the initial parameterisation of some prior\textemdash the choice of constraints on our prior probability density\textemdash is arbitrary explains why approximate Bayesian inference works for arbitrary choices of prior probability: the choice of a prior is mathematically a free choice of gauge.

\subsection{From gauge symmetry to dynamics}

From this point, we can move from a mechanical theory to dynamics. The coupling of a gauge field to a matter field induces a sense of ``direction'' in the tangent space of the bundle, in that paths go in certain directions under gauge forces. In particular, we can define what are termed ``horizontal'' and ``vertical'' flows in a gauge theory. 

\begin{figure}[htb!]
    \centering
    \includegraphics[width=0.8\textwidth]{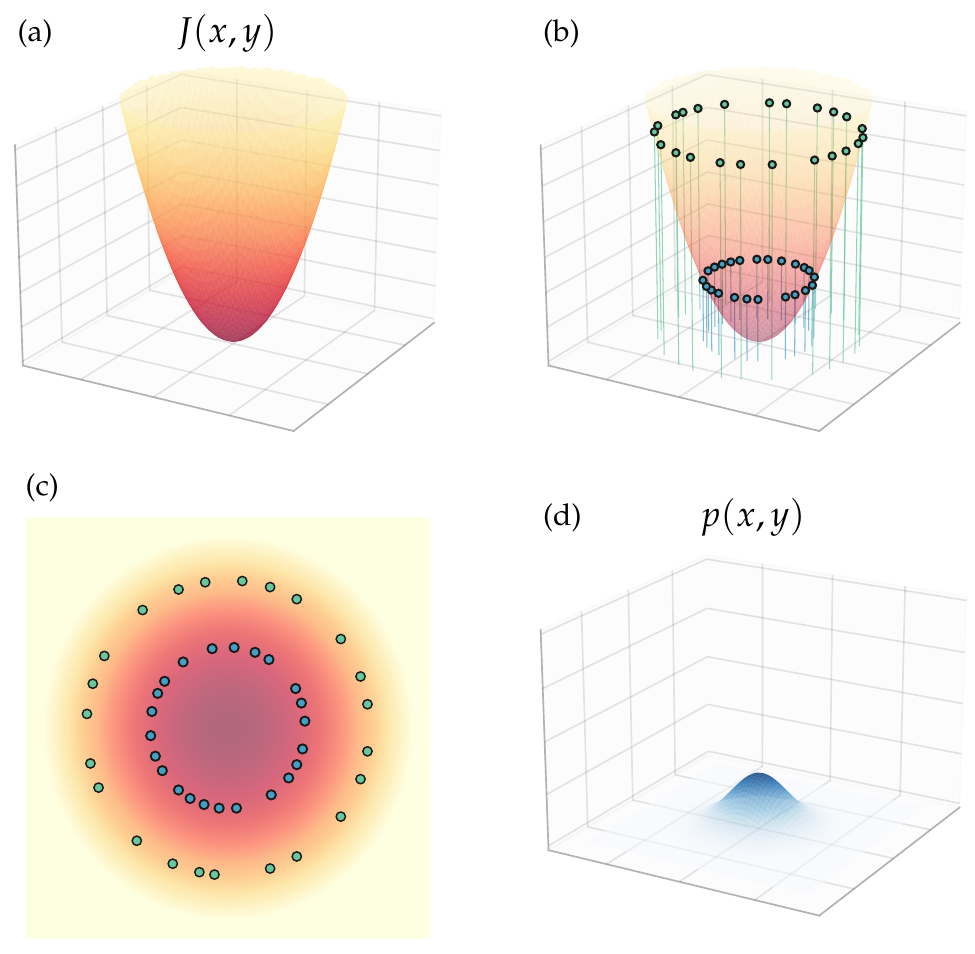}
    \caption{\textbf{A probability density is generated by level sets of the constraint function}. Here it is shown that level sets of $p$ are generated from lifts of level sets of $J$. When flat, the gauge field descends to a vector field (properly a covector field) on the base whose integral curves are level sets $J(x) = c$. Horizontal paths are lifts of these integral curves. They can be defined identically by the parallel transport through the bundle of a lifted point, and yield rings making up $p(x)$.}
    \label{fig:constraint_geom}
\end{figure}

In Figure \ref{fig:constraint_geom}, we introduced the idea of a three-fold structure to gauge theories: a base space $X$ for our space-time or background field, a principal bundle $P$ for choices of gauge over $X$, and an associated bundle $E$ over $X$ coupled to $P$, which tells us how our matter field covaries with the choice of gauge in $P$. When we generate surfaces in the associated bundle $E$, the choice of section underlying that surface is implicitly coupled to the choice of section in the principal bundle $P$, such that changing that choice changes the surface in $E$.

Horizontal paths through a bundle space are very special paths called \textit{horizontal lifts}. These are flat paths, along which there is no change in the choice of gauge\textemdash and hence, no gauge force. Gauge forces deflect horizontal paths by accelerating them, curving them in vertical directions\textemdash hence, changing the choice of gauge as a particle evolves on the base twists its path ``up'' or ``down.'' If this twisting exists for every possible path, the bundle is said to be \emph{curved}, which contrasts with \emph{flat} bundles that admit globally horizontal paths. These are also called globally trivial bundles. Despite the constraining language of all paths being curved, a bundle with even one globally horizontal path is a special construction; in general, fibre bundles are non-trivial (i.e., they feature some curvature).

The object that determines whether a lift\textemdash the image of the generalised function we called a section\textemdash evolves in a flat or curved fashion is a generalisation of a derivative, called a connection, which we discussed at the end of the previous section. A connection induces a vector field along the curve, or conversely, a vector field whose integral curve is the lift (i.e., the image of the section). Under mild assumptions, when a connection is flat everywhere\textemdash contains no vectors with vertical components\textemdash the bundle is trivial. A horizontal vector field like a flat connection is also called a foliation, and foliated vector fields have unique solutions.

We can pull the connection back along the section to get a vector field on the base, called a pullback connection. This determines how particles move on space-time under the influence of gauge forces. The pullback of the connection is what we call a ``local gauge field,'' which is the determinant of this motion. 

The integral curves of the local gauge field are isocontours of the constraint function. In Figure \ref{fig:constraint_geom}, we denoted these as circular level sets of the function $J(x, y) = x^2 + y^2,$ and pulled those circles back to the base. These circles are gauge-horizontal paths: they experience no vertical curvature on the surface $J$.

Feeding this into $E$, we can produce a section of the associated bundle that is gauge-horizontal under an induced connection in $E$. We want horizontality in the constraint space to translate into equiprobability\textemdash gauge-horizontal paths ought to be lifted such that they are also horizontal in $E$, and thus, are rings of equiprobable states. In other words, this is a request that the rings of probability in $E$ are parallel to the rings pulled back to $X$. 

The name for this is \textit{parallel transport}. Parallel transport generates horizontal lifts, in the sense that lifting a point at the start of a path in the base and transporting it parallel to that path generates a horizontally lifted path. As such, we really seek to prove the following: the observation that the shape of a probability density is constrained by the constraints can be made precise in the sense that probability transports itself across the state space in parallel fashion with respect to the constraints.

Equationally, we can derive a striking result: the solution to maximum entropy \emph{is} the parallel transport equation. The function $p$ which maximises entropy is in general $\exp{-\lambda J(x)}$ for some constraint function $\lambda J(x)$. The condition that a single internal function within a section (and hence a point, in the sense of a function evaluated on an input) is translated parallelly can be expressed as an ordinary differential equation (ODE), 
\[
\dd{p(x)} = -\lambda \dd{J(x)} p(x)
\]
which becomes the more familiar first-order ODE
\begin{equation}\label{par-trans}
\pdv{x} p(x) = -\lambda \pdv{x} J(x) p(x)
\end{equation}
in the standard basis of $\R^n$. Note that \eqref{par-trans} rearranges to 
\[
\pdv{x} \log p(x) = - \lambda \pdv{x} J(x)
\]
when dividing both sides by $p(x)$. This equation integrates to \eqref{max-ent-root}, and was our previous equation for motion constrained to lie in the gradient of a potential (or, correspondingly, a connection). Indeed, the solution to the parallel transport ODE \eqref{par-trans} is the exponential function,
\[
p(x) = e^{-\lambda J(x)}.
\]
This proves that the maximum of entropy is parallel transport over the state space, in the same sense as parallel transport is a variational principle for the possible motion in a state space. These probabilistic geodesics are equiprobable rings that comprise the desired probability density.

\subsection{Splitting the flow}

What is the use of discussing the splitting of the flow into horizontal and vertical components in this gauge-theoretic sense? The answer is that doing so provides a natural home for results in the FEP, which clarifies the formal structure of the Helmholtz decomposition; see equation (3). We expand upon the striking result: that the splitting of the flow of a system into vertical and horizontal componenets under the CMEP is isomorphic to the Helmholtz decomposition of the flow of the autonomous partition of a particular system (see \cite{Sakthivadivel2022b} for more a comprehensive technical discussion).

Recall that the deterministic part of the flow of a particular system (the drift component of its SDE) can be decomposed into a solenoidal, probability-mass conserving component, which circulates on isoprobability contours of the NESS density, and a dissipative component which counters random fluctuations. Since horizontal flow is equiprobable, any horizontal flow does not change the value of the surprisal on the states visited. This has been identified as an \textit{exploratory component} of flow in the FEP. As such, if we are modelling organised but itinerant systems, such as those that exhibit life-like characteristics, we can reproduce that tendency to explore formally, by specifying a horizontal component to the flow in this constraint geometry. Inversely, if we consider a very simple system, like a system with linear response\textemdash a system that dissipates, which we expect to be highly constrained about its fixed point\textemdash we can place a very narrow density around that point and build into our model the fact that the horizontal flow should degenerate. 

Conversely, with a privileged horizontal flow, we have a corresponding idea of \textit{vertical flows}. (The horizontal flow is privileged in the sense that it corresponds to inertial paths through state space, i.e., those which are not subject to any extrinsic forces.) Since the vertical flows are identified with gauge forces accelerating the paths out of the horizontal plane, this vertical flow goes to the point of maximum probability\textemdash i.e., the mode of $p(x)$.

It should be noted that this construction assumes that the constraints\textemdash and thus the mode\textemdash are fixed, but, that we can re-maximise entropy and re-fix the gauge anytime we need to update our beliefs about the system. Indeed, the gauge-theoretic view that we introduced here is precisely that this is possible \emph{because} $p(x)$ is covariant on $J(x)$. Future work should extend this to non-stationary regimes; and has, to a limited extent, already begun \cite{barp_geometric_2022}, where the mode and corresponding vertical flow changes direction in time, introducing a continuous interpretation of this iterated inference. Indeed, the suggestion of a continuous view of marginal beliefs along a path can be derived from the principle of maximum path entropy or \textit{maximum calibre} \cite{dill}, which we have previously conjectured is an attractive foundation for extensions of the technology of the FEP to genuine non-equilibria \cite{paths-comment}. We will explore the extension of the duality between the FEP and CMEP to paths in Section \ref{sec:conclusion}, where we introduce $G$-theory. However, we leave details to future work.

To recapitulate\textemdash why have we introduced the technology of gauge theory to Bayesian mechanics? It offers an attractive formulation of approximate Bayesian inference, and is consistent with the manner in which mechanical theories are usually written in contemporary physics; but it is mainly useful because of what it allows us to say about self-evidencing. Previous work \cite{Sakthivadivel2022b} has used the CMEP to prove the approximate Bayesian inference lemma, and tied this result in with a gauge symmetry in CMEP. With this relationship at hand, is absolutely natural to show that the sort of mode-tracking in the approximate Bayesian inference lemma exists, in that realisations of a system under that density flow towards a mode. The splitting evinced by the gauge symmetry allows us to define the Helmholtz decomposition by definition of approximate Bayesian inference.

Given what we have just reviewed, the gauge-theoretic relationship between constraints and probabilities is more useful still, because we can understand now: (i) why surprisal or negative log-probability is the canonical choice of functional, via parallel transport, (ii) why parallel transport makes sense, via gauge covariance identities, and (iii) why there appears to be a force\textemdash a sort of metaphorical life-force, perhaps, and in fact, a gauge force\textemdash driving the mode-matching that underlies the control of probabilistic (Bayesian) beliefs under the FEP.

\subsection{The duality of the FEP and the CMEP}

Despite not being about non-equilibrium systems explicitly,\footnote{Although, note that any directed-ness to the horizontal flow breaks detailed balance, allowing us to extend the CMEP to NESS densities. Also, note that the information-theoretic notion of inferring a stationary probability density is valid even outside equilibrium, whether the system is maximising physical entropy or not.} the CMEP formulation allows us to derive insights from the technology of the FEP. We now review some consequences of the duality. 

Previously, we said that duals are precise opposites: they give two opposing perspectives on a situation; here, one faced inwardly, from the heat bath towards the self-organizing particle, and one faced outwardly. The key is this: dual pairs tell the same story from opposing points of view. Thus, particles that minimise their free energy given some generative model can be understood equivalently as maximising their entropy under a particular constraint. As such, we have a duality of \textit{action functionals} in each description. This is as simple as negating the free energy, converting the log-probability to a constraint, and then maximising the result instead of minimising it, as discussed in \cite{Sakthivadivel2022b}; more generally, it can be obtained by Legendre-Fenchel duality, and our construction (subtracting an entropy off of an internal energy to get a free energy) is loosely equivalent to that duality in this case. A further duality is at hand\textemdash we can interchange internal and external states owing to the symmetry of the Markov blanket, and introduce maximising entropy under a constraint as the dual statement about a system's assumed identity (i.e., expressing the particular system in terms of constraints over the flow of its states). These two adjunctions, relating to the action functional and to the shape of the flow in state space, mean that we can write down mechanical theories for particular systems under some ontological potential or constraint in two dual ways: as a \textit{NESS density}, or as the \textit{section of a principal bundle}. Maximising self-entropy\textemdash \textit{given some specific constraints}\textemdash is equivalent to (in the very precise sense of being dual to) minimising variational free energy \textit{given some generative model}. Constraints subsume system-ness under the CMEP; and geometrically, they are dual to the NESS potential, which plays the same core role. In turn, constraints serve as a potential for the diffusion equation given by a gradient ascent on entropy; just as the NESS density serves as a potential for the flow given by descent on free energy. Finally, we have seen that constraints shape the dynamics of an inferential process in the same way as a gauge field does when it interacts with a matter field (i.e., allowing for redundancy and some degrees of freedom with respect to which the action remains invariant or stationary).

We thus have discussed a mathematically more familiar formulation of the FEP, re-deriving FEP-theoretic results under a complementary principle of stationary action (i.e., the CMEP). A direct consequence of the above is that self-organisation, as described by the FEP, occurs \textit{because of} entropic dissipation, \textit{not} despite it. The idea of maximising entropy allows us to say that life is actually statistically \emph{favoured} as a vehicle for the second law. In other words, despite the apparent paradox of self-organisation in the face of entropy, organisation is encouraged by the universe because order in one place means greater disorder in another (for applications to the FEP, see the argument here: \cite{Ueltzhoffer2020}).

These constraints can be formulated in terms of a Markov blanket in the \textit{state space} of a particular system, such that they are equivalent to a generative model; but there are distinct benefits to the formulation as constraints over what one might call the \textit{existential variables} \cite{Andrews2021} of a system. At one very zoomed out level, having a particular partition leads to the constraint that internal states model external states; so, having a blanket implies certain constraints, and conversely, we can constrain internal states to model external states. However, this can be translated into an equivalent construction that emphasises the way that the internal states of a particular system must look, i.e., the way they are constrained to be optimal parameters of some belief, which we have referred to as an ontological potential. As such, the CMEP formulation allows us to avoid (at least some) of the philosophical concerns that one might have with Markov blankets in state space (e.g. \cite{Raja2021}), at the expense of being less computationally tractable: specifying a full set of existential variables for a system is in general a difficult problem.

The duality of the FEP and the CMEP also sheds light on how to interpret the two components of the flow of autonomous (active and internal) states of a system that arise when we assume that the mechanics of a system has a NESS solution. We have seen that, given a NESS density, the self-evidencing dynamics of a particular system has two components: an dissipative, curl-free component and a solenoidal, gradient-free component, both of which together determine the inferential dynamics of a system. From the perspective of the FEP, this dissipative component can be thought of as a ``fast" flow \textit{towards} the synchronisation manifold \cite{friston2022free} (i.e., one that counters the fast random fluctuations); equivalently, from the dual perspective of the CMEP, this can be cast as a vertical lift into a probability space \cite{Sakthivadivel2022b}; see also Section \ref{sec:math_prelim}. The orthogonal, solenoidal component of the autonomous flow that can be seen as a slow flow \textit{on} the synchronization manifold. 

From an information-theoretic perspective, one can also think of the solenoidal flow as the prediction component of inference, and the dissipative flow as the update term, which corrects the prediction given sensory information \cite{friston2022free}. The solenoidal component of the flow is the estimation part of the inference: it circulates along what one can think of equivalently as the isocontours of a NESS potential, or the level sets of a maximum entropy probability density. In the gauge-theoretic formulation, a horizontal flow builds a piece of the probability density along some contour line. It is these horizontal flows that correspond to the paths of least action that underwrite the FEP, in the absence of random fluctuations. Maximum entropy inference corresponds to this horizontal movement on a statistical manifold, which is often cast as a kind of inference in the FEP literature, in that it circles around the contours of the solution, yielding a whole posterior over state space. But this is not quite inference in the sense of the updating of probabilities. The orthogonal, dissipative flow can be read as the component of the flow that ``corrects" the solenoidal flow, given perturbations induced by sensory states. It thus carries information about external states that is vicariously encoded by sensory states. Dissipative flow is more straightforwardly associated with inference, and is closely associated with updating one's predictions based on sensory data. It ultimately underwrites the existence of the synchronisation manifold between expected internal states (or internal modes) and expected external states (or external modes), and its existence is really the core of the FEP. Indeed, the solenoidal component is not explicitly needed to get Bayesian mechanics\textemdash but it can be specified to get inferential dynamics from particular systems; see \cite{Sakthivadivel2022b, Sakthivadivel2022a}.
\section{The philosophy of Bayesian mechanics}\label{sec:discussion}

The aim of this paper has been to formally introduce the field of Bayesian mechanics as well as its core results and technologies. In this section, we leverage the previous discussion to clarify some of its core philosophical commitments. We provide the following caveat: the philosophy of Bayesian mechanics is obviously a work in progress; we just sketch some crucial points here.

\subsection{Clarifying the epistemic status of the FEP}

We hope that our formal treatment of Bayesian mechanics clarifies issues about what kind of thing the FEP is, and what its epistemological status might be\textemdash as well as its relation to the other main characters of the Bayesian mechanical story (especially the CMEP). The FEP has sometimes been presented using the word ``theory,'' which might imply that is an \textit{empirical theory}\textemdash i.e., that it is the kind of thing that is susceptible to direct confirmation or disconfirmation. Over the nearly two decades since its introduction into the literature as a formal object of discussion\textemdash where it was presented initially as a theory (e.g., a ``theory of cortical function'' or a ``unified brain theory'' in \cite{Friston2005, Friston2010})\textemdash a whole body of work has emerged around it. The FEP and FEP-adjacent work have also been discussed less as a theory, and more as a broad modelling framework (the ``active inference framework''; e.g., \cite{Millidge2020}), or as modelling heuristic (a kind of ``trick,'' in a complimentary sense akin to the ``variational autoencoder trick'' in machine learning, \cite{Raja2021}). The FEP has sometimes been presented as a new branch of physics (a ``particular physics,'' \cite{Friston2019}; or a ``physics of sentient systems,'' \cite{Ramstead2019}), or a self-correcting mathematical approach to typologizing of kinds of systems that exist physically (a ``formal ontology,'' \cite{Ramstead2021, Ramstead2018}). All these perspectives can be reconciled and made sense of. Recently \cite{Andrews2021} usefully used the resources of the philosophy of scientific modelling to draw out attention to the fact that the FEP itself is what we called in Section \ref{sec:preliminaries} a mathematical theory: a formal structure without specific empirical content (i.e., that has no specific empirical application). Our contribution to this discussion has been to draw attention to the role of different kinds of formal structures within Bayesian mechanics.

To sum up: principles like the FEP, the CMEP, Noether's theorem, and the principle of stationary action are mathematical structures that we can use to develop mechanical theories (which are also mathematical structures) that model the dynamics of various classes of physical systems (which are \textit{also} mathematical structures). That is, we use them to derive the mechanics of a system (a set of equations of motion); which, in turn, are used to derive or explain dynamics. A principle is thus a piece of mathematical reasoning, which can be developed into a method; that is, it can applied methodically\textemdash and more or less fruitfully\textemdash to specific situations. Scientists use these principles to provide an interpretation of these mechanical theories. If mechanics explain \textit{what} a system is doing, in terms of systems of equations of movement, principles explain \textit{why}. From there, scientists leverage mechanical theories for specific applications. In most practical applications (e.g., in experimental settings), they are used to make sense of a specific set of empirical phenomena (in particular, to explain empirically what we have called their dynamics). And when so applied, mechanical theories become \textit{empirical theories} in the ordinary sense: specific aspects of the formalism (e.g., the parameters and updates of some model) are systematically related to some target empirical phenomena of interest. So, mechanical theories can be subjected to experimental verification by giving the components specific empirical interpretation. Real experimental verification of theories, in turn, is more about evaluating the evidence that some data set provides for some models, than it is about falsifying any specific model \textit{per se}. Moreover, the fact that the mechanical theories and principles of physics can be used to say something interesting about real physical systems at all\textemdash indeed, the striking \textit{empirical fact} that all physical systems appear to conform to the mechanical theories derived from these principles; see, e.g., \cite{isomura2022experimental}\textemdash is distinct from the mathematical ``truth" (i.e., consistency) of these principles. 

\subsection{\'Elan vital and the FEP}

The idea of an ontological potential endows even simple physical systems, such as rocks, with a kind of weak coherence and ``monitoring'' of internal states (see the typology of kinds of particles in \cite{path-integrals}). Hence, the FEP itself has nothing instructive to say of the demarcations defining life or consciousness \emph{per se}. We have spoken at length about how the viewpoint of constrained entropy makes it apparent how general the FEP is. Namely, the FEP covers a broad class of objects as cases of particular systems, including adaptive complex systems like human beings, simpler but still complex systems like morphogenetic structures and Turing patterns, and even utterly simple, inert structures at equilibrium, like stones on some permissible time-scale. Objects that have no structure or no environment, either of which fail the FEP for obvious reasons, exist at one extreme\textemdash but to make any conclusive statements about the distinctions between living and non-living, or conscious and non-conscious, systems should be regarded as \textit{impossible} in the framework suggested here, and for principled reasons. 

On the other hand, the well-definiteness of adaptive systems that exhibit control but that we would not ordinarily describe as ``cognitive'' has been discussed quite recently under the rubric of ``scale-free cognition'' in \cite{levin-TAME, RN1229}, where it is argued that anything that is an organised system generated via emergent dynamics does indeed satisfy some core properties of cognition. In other words, patterns can be taken as performing inference ``over themselves.'' The consequence that we cannot construct a useful demarcation between \emph{bona fide} cognition and dynamics appearing merely ``as if'' they are cognitive, but which actually reduce to ``mere'' physics, encapsulates the principle of free energy minimisation in the context of cognition in unconventional substrates. These normative statements that everything which could be modelled as performing some kind of inference \textit{can indeed} be understood as performing an elemental sort of inference (as a species of generalised synchrony), without the metaphysical baggage of statements about ``mind" and ``cognition," are restorative. 

On this basis, we can ask whether the FEP really loses some explanatory power as a result of being vacuously true for all sorts of particles. Having originated in the study of the brain, it might seem dissatisfying that the FEP should also extend to inert things like stones, and that its foundations have nothing unique to say about the brain (or the mind, or living systems, for that matter). In our view, the fact that the FEP does not necessarily have anything special to say about cognition is something of a boon\textemdash it should be the case that cognition is like a more ``advanced" or complicated version of other systems, and possesses no special un-physical content. Indeed, the commitment to a principled distinction between cognitive and non-cognitive systems, or living and non-living ones, commits to a sort of \emph{\'elan vital}, wherein the substance and laws of learning, perception, and action should not be grounded in the same laws of physics as a stone, as though they provide a different, more implacable sort of organisation or coherence of states \cite{vital1}. In fact, the opposite has been argued in this paper: that such a theory should be reinterpreted in thermodynamical terms, just as much of the rest of soft matter and biological physics \cite{England2015, england2, Ramstead2018, vital2}. As such, we reject these implicitly dualistic views. As has been suggested many times in these results, the vacuousness of the FEP is really a consequence of its generality, and \emph{that} allows us to look at any system and ask what the FEP says about how we can understand its dynamics. This moves us towards a genuine teleology of self-organising systems \cite{nahas}, via a mechanistic understanding of how that self-organisation rests on\textemdash and is captured by\textemdash Bayesian beliefs.

\subsection{On maps and territories}

This notion of a ``being a model'' is key to Bayesian mechanics and, especially, to the FEP (which could have been called the ``model evidence principle''). Heuristically, the FEP says that if a particular system exists as a cohesive locus of states over time, then it must entail or instantiate a model of its environment. Some have raised the concern that the FEP conflates the metaphorical ``map" and ``territory" (see \cite{ramstead2022map} for a discussion). The question is: Is the FEP \textit{itself} a probabilistic model (or metaphorical ``map") of self-organizing systems; or does it entail that self-organising systems themselves are, carry, or entail a probabilistic model. In other words, are the models in question being deployed by scientists, or by self-organising systems? 

This worry can be addressed simply by noting that there are two ways in which we say that particular systems can be construed as statistical models under the FEP, which pertain to the two core probability density functions that figure in the formulation: these are the generative model and the variational density \cite{RN1250, Ramstead2019enactive, ramstead2022map}. In a first sense, for a particular system ``to be a model'' is shorthand: it means that the system entails or instantiates the statistical relations that are harnessed in the generative model. As we have seen, the generative model is really just the potential function or Lagrangian (if over paths) of a particular system. It is a mathematical construct that can receive a specific empirical interpretation, as a representation of the full, joint dynamics of a particular system: i.e, a mechanical theory coupling autonomous states or paths to external states or paths. Thus, on one first reading, that a particular system \textit{is} a model of its environment means that we can think of the system itself and its embedding environment as entailing the relationships that figure in a generative model. Since they are cast as instantiating these relationships, we can say that the system ``is" the generative model, being careful to note that this is shorthand, and to not conflate the metaphorical ``map" (our scientific model) and the ``territory" (the target system)---at least at this level of analysis. 

But there is also a second sense of ``being a model'' at play in Bayesian mechanics, perhaps the most important of the two, which licences a stronger \textit{representationalist} interpretation. In some sense, the FEP is a ``map'' (a scientific model, indeed, a probabilistic model) of that particular part of the ``territory" that behaves ``as if it were a map" \cite{ramstead2022map, Ramstead2019enactive}. Rather than being a case of reification, as some have suggested (e.g., \cite{RN1216}), one might instead say that the FEP deploys two nested levels of modelling: that of the scientist or observer, and that of the self-organising system being observed. As we have said, given a particular partition, we can interpret the internal states or paths of a particular system as encoding the parameters of a (variational) density over external states. Thus, under the FEP, we can model the states of a system as if they were encoding beliefs ($q$) about some things to which it is coupled ($p$). As Alex B Kiefer once put it (personal communication, 2021), according the FEP, the best scientific model of self-organising systems is one that models them \textit{as statistical models of their embedding environment}. In this sense, the FEP starts from a radical take on the \emph{nouvelle AI} conception that a brain-body-environment system is its own best model.

\subsection{Blanket-based epistemology}

Finally, in light of this, it should be noted that there is an \textit{implicit epistemology} inherent in Bayesian mechanics, which is evinced by both the FEP and the CMEP \cite{ramstead2022map}. The FEP is a metrological (i.e., measurement-theoretic) statement at its core: it entails that the existence of self-organising systems in our physical universe is (or can be modelled as) a kind of measurement \cite{Friston2019}. And at its core, measurement is a kind of inference. The FEP can be stated heuristically as the claim that to exist is to continuously generate evidence for one’s existence\textemdash what we have called self-evidencing \cite{RN92}. Continued existence in the physical universe provides systems with sensory evidence of their own existence; and under the FEP, self-organisation occurs because systems minimise free energy (i.e., minimise the discrepancy between expected data, given a model of how that data was generated, and sensed data). 

The FEP further implies that, ultimately, there is no way to meaningfully distinguish between saying that the dynamics of system actually engage in or instantiate approximate Bayesian inference, and saying that they merely ``look as if'' they do so\textemdash without, that is, breaking the blanket itself. This is true both from the point of view of scientists measuring self-organising systems and from the point of view of self-organisation itself\textemdash and it is further illuminated by the duality between the FEP and the CMEP. The key point is that measurement is not a given: after all, a measurement is an inference based on data and prior beliefs. This speaks to the observational foundations of physics \cite{cook1994observational, bridgman1954remarks}. Without this ability to measure and infer, arguably, there would simply be no physics. The FEP in some sense captures our epistemic predicament as needing to make inferences from data to learn about the world: it mandates that we can never go ``beyond the blanket.'' As scientists attempting to make sense of some phenomena, we can never escape having to merely \textit{infer} the states of affairs behind the blanket (i.e., the internal states or paths of phenomena, given the different kinds of data at hand). In FEP-theoretic terms, from the perspective of our own particular partitions as scientists, we only have access to our sensory paths, i.e., our measurements, which we use to make model-based inferences about what generated our data. And (at least arguably), saying anything about what lies beyond the blanket that would in some sense escape our vicarious relation to it would be more akin to metaphysics than it would be to scientific inquiry. 

Crucially, this kind of blanket-based epistemic minimalism is fully consistent with (indeed, dual to) the implications of Jaynes' maximum entropy principle, which is precisely about modelling physical systems from a point of view of maximal ignorance. The maximum entropy principle is used to fashion a probability density to explain some data based on as few assumptions as possible. The maximum entropy principle says that given a set of distributions that might explain some data, the one with the most entropy (i.e., the most uninformative one) is the ``true'' distribution. Likewise and dually, and more formally, the FEP says that given a set of sensory states or paths (i.e., given some data), the real path of autonomous states is the one expected have the least free energy (i.e, the least surprising one). The philosophical duality also expresses itself practically. Using the technology of the CMEP, we can create mathematical models of a particular system, where we occupy the perspective of the environmental system or external observer that is embedding the particular system. Dually, using the FEP, we can also model how a particular system measures itself and its environment from its own perspective \cite{ramstead2022map}. Thus, even philosophically, the FEP and the CMEP are really at their core two sides of the same coin. Thus, we can read the FEP as a physics \textit{of} beliefs, in the sense that it is a principle allowing us to formulate mechanical theories for the image of a particular system in the space of beliefs; and dually, we can read the CMEP as a physics \textit{by} beliefs, in the sense that it is a principle specifying how to use the formal structure of probabilistic belief updating to model particular systems.
\section{Concluding remarks and future directions: towards \textit{G}-theory}\label{sec:conclusion}

We have presented the current state of the art in Bayesian mechanics. After some preliminary discussion, we reviewed core results of the FEP literature, examining a three-part typology of the kinds of systems to which the FEP has been applied (over paths, over states with a NESS potential, and over states at a stationary NESS). We then reviewed the duality of the FEP and the CMEP. We saw that one can construct a gauge-theoretic formulation of the CMEP, which explains why approximate Bayesian inference works at all, and why everything looks as if it was becoming a model of its embedding environment on average and over time (and why this is like dissipating into it, given some phenotype-congruent constraints on the mechanics of that process). 

We now briefly discuss one core direction of research in Bayesian mechanics, focusing on the construction of a mathematical theory that will extend the duality between the FEP and the CMEP, and beyond. We started with a path-based formulation of the FEP, and we noted that in order to say more meaningful things about the mechanics of such systems, we could formulate things in terms of probability density dynamics over states. We have seen that the states-wise formulation of FEP is dual to the CMEP, which is formulated in terms of states. This body of work\textemdash in tandem with other work in the field, e.g., \cite{fields2021free, fields2022neurons}\textemdash increasingly suggests that existing approaches to the physics of intelligence and adaptivity fit together in a nontrivial way, as if they were parts of an as yet undiscovered whole. Although currently in its early stages, and still inchoate, one can begin to see the shape of a general mechanics for complex self-organising systems emerge from this convergence.

$G$-theory is the name we have given to a larger theory of complex adaptive systems that we do not yet fully understand, but whose existence is strongly suggested by the duality between the FEP and the CMEP explored here. The name $G$-theory is intended as something like an homage to $M$-theory in theoretical physics. The existence of $M$-theory was first conjectured in the 1990s as a unifying treatment of known string theories \cite{witten1995string}; at the time, it was known that specific versions of string theory were dual to each other, and the idea was that there must exist an underlying theory of which all known string theories were aspects or facets. Like $M$-theory, the reader should note that the `$G$' in $G$-theory does not mean anything specific. If pressed, the authors might suggest `gauge,' or maybe `generalised,' or point out that expected free-energy is often denoted \textbf{G}.

One duality that suggests the existence of $G$-theory is the equivalence of the density dynamics formulation of the FEP and the CMEP, which we discussed at length in this paper. Another, which we have begun to investigate of late \cite{classical-physics}, concerns the equivalence between the paths-based formulation of the FEP and maximum calibre, an extension of the formalism of CMEP to the entropies of paths or trajectories. The entropy of an ensemble of paths is known as ``calibre'' \cite{Jaynes1957}; and the formulation of maximum \textit{path entropy} is known as maximum calibre. 

Here, our discussion comes full circle. We started from a discussion of paths of stationary action, then moved from there to density dynamics formulations (i.e., probability densities over states) to see what could be said with some addition assumptions in play. We can now \textit{return to paths}, but from the perspective of the \textit{maximisation of entropy}. In ongoing and future work, we aim to construct relationships of the following form, further dualising the Bayesian-mechanical construction:
\[
\begin{tikzcd}[row sep = 6em, column sep = 4em]
-\int p(x) \log p(x) \dd{x} - \E[\sigma^{-1}(\hat\eta)] + \hat\mu \arrow[d] \arrow{r} &-\int p(\gamma) \log p(\gamma) \dd{\gamma} - \E[\sigma^{-1}(\hat\eta(t))] + \hat\mu(t) \arrow{d}\\
D_{\text{KL}} [q(x) \| p(x)] +\log p(x) \arrow{r} \arrow[u, shift left=0.75em] \arrow[ur, dotted] &D_{\text{KL}} [q(x,t) \| p(x,t)] +\log p(x,t) \arrow[u, shift right=0.75em, swap, "\,\, G\text{-theory}"]
\end{tikzcd}
\]
The ``adjunction'' between constrained self-entropy and free energy on beliefs discussed here and in \cite{Sakthivadivel2022b} is the left-most pair of maps on the diagram. The map on the top, generalising maximum entropy to maximum calibre, has been discussed in \cite{ying-jen, dill}; and the relationship between free energy and expected free energy has been worked out previously in several places, e.g., in \cite{path-integrals, gefe, efe}, as we discussed in Section \ref{sec:sentientsystems}. One aspect of $G$-theory is represented in the above diagram. A first case of $G$-theory would consist of a map on the right,
\[
D_{\text{KL}} [q(x,t) \| p(x,t)] +\log p(x,t) \longrightarrow -\int p(\gamma) \log p(\gamma) \dd{\gamma} - \E[\sigma^{-1}(\hat\eta(t))] + \hat\mu(t),
\]
along with its adjoint, constructed such that the diagram commutes in the direction of the dotted line (a map whose existence is sketched out in \cite{paths-comment} in particular). This will allow us to further leverage the equivalence between the FEP and the constrained minimisation of entropy. Whilst the implications of this have yet to be fully worked out, these technologies will allow us to write down mechanical theories for the kinds of systems that are more easily expressible in terms of calibre as opposed to surprisal and its variational free energy bound. This may turn out to be a non-negligible class of systems, since many biological systems seem to appear to be non-stationary, to have moving attractors, to have chaotic trajectories \cite{classical-physics}, or to have no steady state density, at least over some timescales. Indeed, as has been remarked \cite{Sakthivadivel2022b}, maximum calibre is perhaps a more natural setting for dealing formally with these kinds of systems systems.

In closing, we hope to have provided a useful introduction to Bayesian mechanics, and clarified the core notions and constructs that it involves. We have seen that Bayesian mechanics comprises tools and technologies that allow us to leverage a principle (i.e., the FEP or the CMEP) to write down mechanical theories for stochastic systems that look as if they are estimating posterior probability distributions over the causes of their sensory states. We have seen that Bayesian mechanics is specialised for systems with a particular partition, which allow the particular system to encode the parameters of probabilistic beliefs about the quantities that characterise the system as ``the kind of system that it is." Bayesian mechanics thereby provides a formal language to model the constraints, forces, fields, manifolds, and potentials that determine how the image of a physical, particular systems moves in a statistical manifold. We reviewed the main FEP-theoretic results in the literature, using a typology of the kinds of systems to which it applies. We also reviewed core notions from the philosophy of scientific modelling as it applies to physics, as well are core constructs from mechanics and gauge theory. We then discussed the duality of the FEP and the CMEP, which lie at the heart of Bayesian mechanics, and examined the deep implications of this duality for Bayesian mechanics and physics more broadly. We are enthusiastic about recent and future developments of Bayesian mechanics. Mathematical formulations for Bayesian mechanics are available\textemdash much remains to be done.
\section*{Additional information}\label{sec:additional_info}

\subsection*{Acknowledgements}\label{sec:ack}
The authors thank Miguel Aguilera, Mel Andrews, Chris Buckley, Chris Fields, James Glazebrook, Harrison Hartle, Alex Kiefer, Alec Tschantz, Kai Ueltzh\"offer, and the members of the VERSES Research Lab for valuable comments on early versions of this work, and for useful discussions that shaped the contents of the paper. We also thank the editors and co-editors of a special issue of the journal \emph{Royal Society Interface}, ``Making and Breaking Symmetries in Mind and Life,'' to which this contribution was invited.

\subsection*{Funding statement}\label{sec:fund}
The authors are grateful to VERSES for supporting open access publication of this paper. CH is supported by the U.S. Office of Naval Research (N00014-19-1-2556). BK \& CH acknowledge the support of a grant from the John Templeton Foundation (61780). BM was funded by Rafal Bogacz with BBSRC grant BB/s006338/1 and MRC grant MC UU 00003/1. LD is supported by the Fonds National de la Recherche, Luxembourg (Project code: 13568875). KF is supported by funding for the Wellcome Centre for Human Neuroimaging (Ref: 205103/Z/16/Z) and a Canada-UK Artificial Intelligence Initiative (Ref: ES/T01279X/1). This publication is based on work partially supported by the EPSRC Centre for Doctoral Training in Mathematics of Random Systems: Analysis, Modelling and Simulation (EP/S023925/1). The opinions expressed in this publication are those of the author(s) and do not necessarily reflect the views of the John Templeton Foundation.

\printbibliography[title={References}]

\end{document}